\documentclass[aps,pre,11pt,byrevtex,a4paper,nofootinbib,showpacs,showkeys,longbibliography]{revtex4-1}
\pdfoutput=1
\usepackage{microtype}
\usepackage{amssymb}
\usepackage{amsmath}
\usepackage{graphics}
\usepackage[colorlinks,linkcolor=blue,urlcolor=blue,citecolor=blue]{hyperref}
\usepackage[applemac]{inputenc}
\usepackage{lineno}

\newcommand{\be}{\begin{equation}}
\newcommand{\ee}{\end{equation}}
\newcommand{\om}{\omega}
\newcommand{\ra}{\rightarrow}

\newcommand{\reals}{\mathbb{R}}
\newcommand{\cL}{\mathcal{L}}

\newcommand{\cM}{\mathcal{M}}
\renewcommand{\Pr}{\mathbb{P}}
\newcommand{\Qr}{\mathbb{Q}}
\newcommand{\Ex}{\mathbb{E}}
\renewcommand{\dag}{\dagger}
\newcommand{\micro}{\textrm{micro}}
\newcommand{\cano}{\textrm{cano}}
\newcommand{\tA}{\tilde A}
\newcommand{\hF}{\hat F}
\newcommand{\p}{\partial}
\newcommand{\inv}{\textrm{inv}}
\newcommand{\Pra}[1]{\overset{#1}{\longrightarrow}}
\newcommand{\hX}{\hat X}
\newcommand{\hpsi}{\hat\psi}
\newcommand{\tu}{\tilde u}
\DeclareMathOperator*{\statpt}{stat.pt.}
\DeclareRobustCommand{\topcons}{\genfrac{}{}{0pt}{}}

\begin{document}

\title{Variational and optimal control representations of conditioned\\ and driven processes}

\author{Rapha\"el Chetrite}
\email{Raphael.Chetrite@unice.fr}
\affiliation{Laboratoire J.~A. Dieudonn\'e, UMR CNRS 7351, Universit\'e de Nice Sophia Antipolis, Nice 06108, France}

\author{Hugo Touchette}
\email{htouchette@sun.ac.za, htouchet@alum.mit.edu}
\affiliation{National Institute for Theoretical Physics (NITheP), Stellenbosch 7600, South Africa}
\affiliation{Department of Physics and Institute of Theoretical Physics, Stellenbosch University, Stellenbosch 7600, South Africa}
\date{\today}

\begin{abstract}
We have shown recently that a Markov process conditioned on rare events involving time-integrated random variables can be described in the long-time limit by an effective Markov process, called the \emph{driven process}, which is given mathematically by a generalization of Doob's $h$-transform. We show here that this driven process can be represented in two other ways: first, as a process satisfying various variational principles involving large deviation functions and relative entropies and, second, as an optimal stochastic control process minimizing a cost function also related to large deviation functions. These interpretations of the driven process generalize and unify many previous results on maximum entropy approaches to nonequilibrium systems, spectral characterizations of positive operators, and control approaches to large deviation theory. They also lead, as briefly discussed, to new methods for analytically or numerically approximating large deviation functions.
\end{abstract}


\maketitle

\tableofcontents

\newpage
\section{Introduction}

We continue in this paper our work on stochastic processes conditioned on large deviations~\cite{chetrite2013,chetrite2014}. The idea is to describe the evolution of a stochastic process when one or more `observables' of this process are observed to fluctuate in time away from their typical values. The approach followed is based on large deviation theory and proceeds by conditioning a Markov process $X_{t}$ on a rare event, defined in terms of a functional $A_{T}$ of the process integrated over the time interval $[0,T]$, and by deriving from this conditioning a new Markov process $\hX_t$ -- called the  \emph{auxiliary}, \emph{effective} or \emph{driven process} -- which can be proved to be equivalent to the conditioned process $X_t|A_T$ as $T\ra\infty$. In this limit, it is thus possible to describe the trajectories, paths or histories of $X_t$ leading to rare fluctuations of $A_T$ by a conditioning-free Markov process identified as the driven process.

We have explained in \cite{chetrite2013,chetrite2014} how this conditioning problem relates in physics to generalizations of the microcanonical and canonical ensembles defined on paths of nonequilibrium systems, and how the driven process is constructed mathematically by a generalization of \emph{Doob's $h$-transform}, used in probability theory to describe simple conditionings of Markov processes \cite{doob1984,doob1957,rogers2000,darroch1965,darroch1967,collet2014}. Physically, the driven process can also be seen as a generalization of the notion of \emph{fluctuation paths} describing dynamical fluctuations in low-noise systems in terms of most-likelihood paths (also called \emph{escape} or \emph{reactive paths}). The theory of these paths, widely used in physics, biology and chemistry to describe noise-activated processes \cite{kampen1992,gardiner1985,hanggi1990,melnikov1991}, goes back to Onsager and Machlup \cite{onsager1953} and has come to be formalized in the 1970-80s as the Freidlin-Wentzell-Graham (FWG) large deviation theory of dynamical systems perturbed by noise \cite{freidlin1984,graham1989,luchinsky1998,touchette2009}. Our approach generalizes this theory in that it is not restricted to low-noise or low-temperature systems: it can be applied in principle to any Markov system driven arbitrarily far from equilibrium by external forces, boundary reservoirs, and noise sources to describe, by means of an effective stochastic process rather than a single deterministic path, how fluctuations arise in time.

Other links between the driven process, control theory, rare event simulations, and the physics of nonequilibrium systems are mentioned in \cite{chetrite2014}. Our goal here is to provide more details about the link with control theory, and to show in particular that the driven process can be interpreted as an optimal stochastic control process minimizing a cost function related to the large deviations of the conditioning observable $A_T$. We also show, as a prelude to this control result, that the driven process can be characterized by a number of equivalent variational principles involving large deviation functions and relative entropies.

These principles follow, as most variational principles of statistical mechanics~\cite{ellis1985,oono1989,touchette2009}, from the so-called \emph{contraction principle} of large deviation theory and describe the fact that nonequilibrium systems `build up' fluctuations in an optimal way to reach states far from their typical states. This is very much in the spirit of the Onsager-Machlup principle of minimal dissipation \cite{onsager1953} and of generalizations of this principle forming the basis of the FWG theory. The crucial difference again is that the principles that we derive can be applied to weak- and strong-noise systems, in addition to equilibrium and nonequilibrium systems. 

As in our previous work \cite{chetrite2013,chetrite2014}, the results that we discuss relate to and contain many results obtained before, but also clarify, unify, and generalize them, we believe, in many ways. We briefly discuss these relations and generalizations next; more specific references and explanations will be given in the following sections, especially in Sec.~\ref{seccont} and the conclusion section~\ref{seccon}.

\subsection{Control approaches to large deviations} 
Our main source for this paper is the work of Wendell H. Fleming and his collaborators \cite{fleming1971,fleming1978,fleming1983,sheu1984,fleming1985b,fleming1987,sheu1991,fleming1992b,fleming1992,fleming1995,fleming1997} on transforming linear partial differential equations (PDEs) arising in large deviation theory into nonlinear PDEs that have the form of Hamilton-Jacobi-Bellman (HJB) equations arising in control theory; see also \cite{whittle1990,whittle1991,whittle1994}. From this control mapping, which is essentially a \emph{Hopf-Cole transform} referred to as a \emph{logarithmic transform} by Fleming, one is able to obtain large deviation functions by solving HJB equations using control methods. This can be applied either in the low-noise or the long-time limit of large deviations, the later being the subject of the Donsker-Varadhan (DV) theory \cite{donsker1975,donsker1975a,donsker1976}. 

We have explained this approach briefly in \cite{chetrite2014}; see also \cite{fleming1982,fleming1985,fleming1989} for reviews and \cite{hartmann2012} for another useful summary. The connection with our work comes from the fact that the optimal control process obtained in this approach corresponds to the driven process in the case of time-integrated observables studied in the long-time DV limit. To our knowledge this connection was not made before. It is known from the work of Fleming and Sheu \cite{sheu1984,fleming1985b,fleming1987} that the controlled process corresponds to a change of measure of the original process considered, but no link was established between this change of measure, the generalized Doob transform, and the conditioning problem. 

These connections are established in Sec.~\ref{seccont} and complement Fleming's theory by showing that the optimal process solving a large class of stochastic HJB equations with quadratic cost corresponds to a large deviation conditioning of a non-controlled process. This interpretation applies to low-noise (FWG) large deviation problems, but also to long-time (DV) large deviation problems for systems that are not necessarily perturbed by a weak noise. In addition, we consider, as explained next, the control mapping for a significantly wider class of control costs and observables suggested by nonequilibrium systems.

\subsection{Stochastic control with current-type costs}
 
Historically, optimal control theory and dynamic programming \cite{bellman1967,fleming1975,stengel1994,fleming2006,schulz2006} have been developed for cost functionals of deterministic and stochastic systems having the form
\be
C_t^T=\int_t^T \psi(X_s) ds + \phi(X_T),
\ee
where $\psi$ and $\phi$ are arbitrary functions of the state variable $X_s$ and $t<T$. Applications in nonequilibrium statistical physics require that we consider more general costs that involve not only an integral of $X_s$, but also a sum that depends on the jumps of $X_s$, in the case of a jump process, or an integral over its increments, in the case of a pure diffusion. 

A generalization of stochastic control theory to these costs, which are related physically to particle and energy currents, has been proposed recently by Bierkens, Chernyak, Chertkov and Kappen \cite{bierkens2013,chernyak2014}. Their approach follows that of Kappen \cite{kappen2005b,kappen2005,kappen2011} which is itself a reversal of Fleming's approach: they derive the HJB equation for the solution of a special stochastic optimization problem involving a quadratic cost and then apply a logarithmic transform to this equation to obtain a linear PDE which can be solved using spectral or path integral methods. 

The present paper can be seen as an alternative approach to this generalization of stochastic control and optimization, providing new proofs of the HJB equation for current-type costs. In fact, we provide in Appendix~\ref{apphjb} a simple proof of this equation, which relies only on It\=o's calculus. Our results also provide, as a complement to Fleming's approach, a probabilistic interpretation of the optimal control process and relate quadratic costs to large deviations. This is the subject of Sec.~\ref{seccont}.

\subsection{Spectral characterizations of positive operators}
 
An interesting consequence of the control approach to large deviations is that it provides an interpretation of some variational characterizations of the dominant eigenvalue of linear positive operators~--~in particular, the Rayleigh-Ritz variational principle of quantum mechanics \cite{courant2004}, as well as generalizations of this principle obtained for non-hermitian operators by Donsker and Varadhan \cite{donsker1975b,donsker1975c,donsker1976b} and by Holland \cite{holland1978,holland1977}. A control approach to dominant eigenvalues was also proposed by Fleming and Sheu for general additive costs~\cite{sheu1984,fleming1987,fleming1997} (see \cite{fleming1985} for a review) and is mentioned in the context of current-type costs by Bierkens, Chernyak, Chertkov and Kappen~\cite{bierkens2013,chernyak2014}. 

Our work can be seen as yet another approach to this problem in which the DV characterization of dominant eigenvalues follows as a special case of more general variational principles that we derive from the contraction principle as applied to the so-called \emph{level 2.5 of large deviations}~\cite{maes2008,maes2008a,chernyak2009,bertini2012,barato2015}. The optimizer of these variational principles is the driven process, as will be shown in Sec.~\ref{secvarrep}, so that these principles can be used to characterize that process independently of its relationship with the optimal control process and the conditioning problem.

\subsection{Variational methods for computing large deviations}

The DV characterization of dominant eigenvalues is useful not only from a spectral point of view, but also as a computing tool for obtaining large deviation functions. One important function of large deviation theory, the \emph{scaled cumulant generating function}, is known to correspond for many observables of Markov processes of interest to the dominant eigenvalue of a positive linear operator called the \emph{tilted generator}. Having a variational representation for this eigenvalue allows one to obtain useful approximations and bounds to the scaled cumulant generating function.

Variational principles also exist for another large deviation function, the \emph{rate function}, which is the main function of interest in large deviation theory. The DV variational formula for the level-2 rate function \cite{donsker1975,donsker1975a,donsker1976} is one such principle, as is a variant of that principle derived by Baldi and Piccioni \cite{baldi1999} for finite-space jump processes via the level 2.5 of large deviations. In physics, variational principles for large deviation functions have also been proposed for DV large deviation problems by Eyink \cite{eyink1996a,alexander1997,eyink1998}, who refers to them as \emph{action principles} related to the Rayleigh-Ritz method, by Nemoto and Sasa \cite{nemoto2011,nemoto2011b,nemoto2014} (see also \cite{sughiyama2013}) for special observables of one-dimensional diffusions, and by Jack and Sollich \cite{jack2015} for jump processes.

The results that we derive in Sec.~\ref{secvarrep} include all of these variational principles and show that the optimizer of these principles is the driven process. This provides a clearer understanding of the work of Nemoto and Sasa \cite{nemoto2011,nemoto2011b,nemoto2014} on feedback control methods for estimating large deviation functions. We briefly discuss these methods at the end of the paper in the context of control theory and adaptive importance sampling. 

\subsection{Path maximum entropy and maximum caliber}

A final link exists between our work and the maximum entropy approach to nonequilibrium systems proposed in the 1960s by Filyukov and Karpov~\cite{filyukov1967,filyukov1967b,filyukov1968b}, and re-worked recently by Evans~\cite{evans2004,evans2005a,evans2010} for sheared fluids. The basis of this approach, also known as the \emph{dynamical maximum entropy} or \emph{maximum caliber method} \cite{jaynes1980,stock2008,presse2013}, is to describe the stochastic dynamics of a nonequilibrium process driven in a steady state by a path distribution maximizing the Shannon entropy subject to a constraint  (e.g., current or shear state) describing the steady state. For Markov chains and jump processes, Monthus~\cite{monthus2011} has shown that this maximization yields the driven process, when expressed more generally as a constrained minimization of a path relative entropy. In Sec.~\ref{secvarrep}, we generalize this result to general Markov processes and relate the relative entropy to the level 2.5 of large deviations. This establishes new connections between the dynamical maximum entropy method and all the topics mentioned before, in addition to provide a probabilistic justification of this otherwise \emph{ad hoc} method. 

%

\section{Theory of conditioned and driven processes}
\label{secnot}

We review in this section the construction and physical meaning of the conditioned and driven Markov processes. We follow closely the notations of \cite{chetrite2014}, but restrict ourselves to the case of pure diffusions to simplify and shorten the presentation. Jump processes and Markov chains are considered in Appendices~\ref{appjp} and \ref{appmc}, respectively. Mixed or hybrid processes combining diffusive and jump parts can also be treated using the general language of Markov generators; see \cite{chetrite2014}.

\subsection{Conditioned process}

We consider an ergodic Markov process $X_{t}$ evolving over a time interval $[0,T]$. This process is defined mathematically by its \emph{generator} $L$, whose action on functions $h$ of $X_t$ gives the evolution of their expectation according to
\be
\p_t \Ex_x[h(X_t)]=\Ex_x[Lh(X_t)],
\ee
where $\Ex_x[\cdot]$ denotes the expectation with fixed state $X_t=x$. From this relation, it can be shown that $L$ determines the transition probability kernel
\be
P_s^t(x,y)=e^{(t-s)L}(x,y)
\ee
associated with the transition from $X_s=x$ to $X_t=y$ with $s<t$. Its \emph{dual} $L^\dag$ determines via the \emph{master equation}
\be
\p_t \rho(x,t)=L^\dag \rho(x,t)
\label{eqmeq1}
\ee
the evolution of probability densities (or measures in general). 

Another way to define $X_t$ is to specify its \emph{path measure} $d\Pr_{L,T}(\om)$ which corresponds intuitively to the distribution of its paths $\{X_t(\om)\}_{t=0}^T$. This measure can be defined via finite-dimensional distributions (see \cite{chetrite2014}) and depends on $L$ and $T$, in addition to the initial density $\rho_0(x)=\rho(x,0)$. For simplicity, we assume that all paths start at $X_0=x_0$, so that $\rho_0=\delta_{x_0}$. 

We focus as mentioned on pure diffusions defined by the following (Stratonovich) stochastic differential equation (SDE):
\be
dX_{t}=F(X_{t})dt+\sigma(X_t) \circ dW_{t},\qquad X_{t}\in\reals^{d},
\label{eqsde1}
\ee 
where $F:\reals^d\ra\reals^d$ is the \emph{drift}, $\sigma:\reals^d\ra\reals^d$ is the \emph{diffusion field}, and $W_t\in\reals$ is a Brownian motion.\footnote{See \cite{chetrite2014} for general SDEs involving more than one Brownian motion.} For this model, the generator is given by
\be
L=F\cdot\nabla+\frac{1}{2}(\sigma \cdot \nabla)^{2}=\hF \cdot \nabla+\frac{1}{2}\nabla D \nabla,
\label{eqgen1}
\ee
where
\be
\hF(x)=F(x)-\frac{1}{2} (\nabla \cdot \sigma)(x)\, \sigma(x)
\label{eqFhat}
\ee
is the \emph{modified drift} and $D=\sigma\sigma^T$ is the \emph{covariance matrix}.\footnote{As in \cite{chetrite2014}, we consider the Stratonovich interpretation of the SDE only for convenience; other interpretations can be considered with appropriate changes. For a diffusion matrix $\sigma$ that does not depend on $X_t$, $\hF=F$.} The master equation (\ref{eqmeq1}) is also in this case the \emph{Fokker-Planck equation}, which can be expressed as the continuity equation
\be
\p_t \rho=-\nabla\cdot J_{F,\rho},
\ee
where 
\be
J_{F,\rho}=\hF\rho-\frac{D}{2}\nabla\rho
\label{eqcurr1}
\ee
is the \emph{Fokker-Planck probability current} associated with the drift $F$ and density $\rho$ \cite{risken1996}. The \emph{invariant density} $\rho^\inv_F(x)$ of the process satisfies $L^\dag\rho^\inv_F=0$ or
\be
\nabla\cdot J_{F,\rho^\inv_F}=0
\ee 
and corresponds, under our assumption that $X_t$ is ergodic, to its unique stationary density. 

Physically, we imagine $X_{t}$ to be the state of a stochastic system involving one or many particles that are either at equilibrium or are forced into a nonequilibrium steady state by boundary reservoirs or external forces violating detailed balance. As the system evolves randomly for $t\in[0,T]$, we are interested in tracking a certain observable $A_{T}$, representing, for example, the work done on the system by an external force or the heat exchanged with its environment, and in studying the system's paths leading to \emph{rare} values of $A_T$ that have a low probability of being observed after a long time $T$. This means, following the introduction, that we want to study the behavior of $X_{t}$ \emph{given} that $A_{T}$ is observed to be far from its typical value after a long time $T$.

\begin{figure*}[t]
\centering
\includegraphics{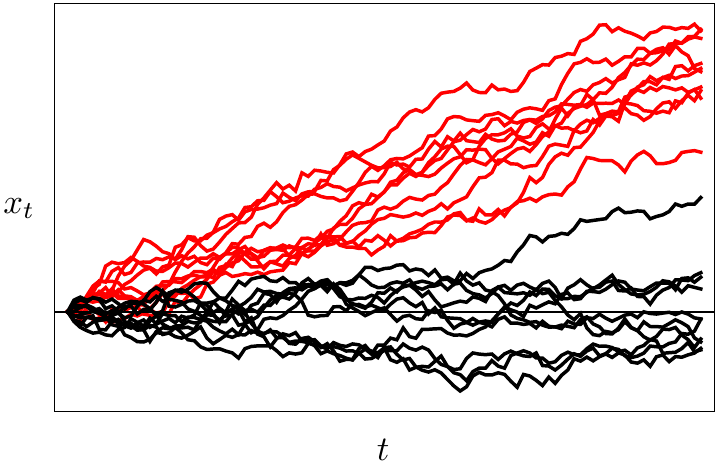}%
\hspace*{0.3in}%
\includegraphics{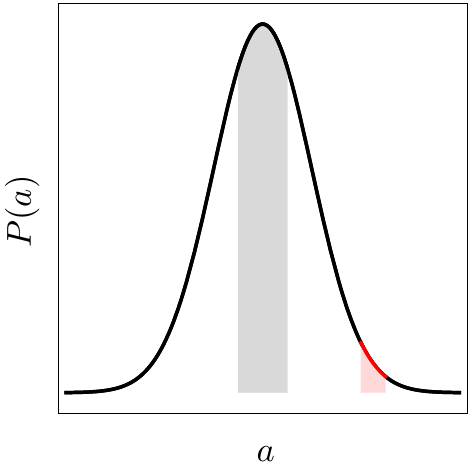}
\caption{Illustration of the conditioned process. Black: Paths of a process $X_{t}$ leading to \emph{typical} values of the observable $A_{T}$ having high probability (gray shaded area). Red: Paths of $X_{t}$ leading to \emph{atypical} values of $A_{T}$ having low probability (red shaded area). Each set of paths defines a conditioned process $X_{t}|A_{T}=a$ characterized by the conditional or microcanonical path measure (\ref{eqmicro1}). The driven process is the homogeneous Markov process describing the conditioned process in the long-time limit $T\ra\infty$.}
\label{fig1}
\end{figure*}

This conditioning of $X_{t}$ is illustrated in Fig.~\ref{fig1} and is defined probabilistically by the \emph{conditional path measure}
\be
d\Pr_{a,T}^{\micro}(\om)\equiv d\Pr_{L,T}\{\om|A_{T}=a\}
=\frac{d\Pr_{L,T}\{\om,A_{T}=a\}}{\Pr_{L,T}\{A_{T}=a\}}.
\label{eqmicro1}
\ee
The superscript `micro' refers to the fact that this conditional measure is a path analog of the microcanonical ensemble representing in equilibrium statistical mechanics the probability measure of a many-particle system conditioned to have a constant energy. In the following, the conditioned process defined by (\ref{eqmicro1}) is denoted by $X_{t}|A_{T}=a$.

An obvious question to ask about $X_{t}|A_{T}=a$ is whether this process is Markovian -- that is, whether it can be described by a Markov generator and, in the case of diffusions, by an SDE that depends on the constraint $A_{T}=a$. To our knowledge, this is not the case for general observables with $T<\infty$ because of the non-local (in time) nature of the constraint $A_{T}=a$, even if we allow for non-homogeneous (i.e., time-dependent) generators. However, as we prove in \cite{chetrite2014}, the conditioned process \emph{does converge in the limit $T\ra\infty$ to a homogeneous Markov process}, corresponding to the \emph{driven process} $\hX_t$ defined in the next subsection. 

This result is derived in \cite{chetrite2014} for a large class of observables $A_{T}$ suggested by physical applications that depend on the state of the process $X_{t}$ and its jumps or increments. For pure diffusions, these observables have the general form
\be
A_{T}=\frac{1}{T}\int_{0}^{T} f(X_{t})dt+\frac{1}{T}\int_{0}^{T} g(X_{t})\circ dX_{t},
\label{eqobs1}
\ee
where $f$ is a scalar function, $g$ is a vector function, and $\circ$ denotes the Stratonovich product.\footnote{The Stratonovich convention is also used here for convenience.} The driven process is also obtained by assuming that the probability distribution of $A_{T}$ satisfies a \emph{large deviation principle} (LDP), which means essentially that
\be
\Pr_{L,T}\{A_T=a\}\approx e^{-TI(a)}
\label{ldpdef1}
\ee
in the limit $T\ra\infty$ with subexponential corrections in $T$. 

This scaling result is found for many observables of nonequilibrium systems~\cite{touchette2009,derrida2007,bertini2007,harris2013} and implies that the probability of $A_T$ decays exponentially with $T$, except at the global minimum and zero $a^*$ of the \emph{rate function} $I(a)$ where it concentrates with $T$ \cite{ellis1985,touchette2009,dembo1998}. Hence, values $a\neq a^*$ represent fluctuations of $A_T$ that are exponentially rare with the observation time $T$, whereas $a^*$ itself represents the \emph{stationary} or \emph{ergodic value} of $A_T$ which becomes most probable as $T\ra\infty$. In mathematical terms, this concentration of probability defines a (weak) law of large numbers, which we express here as 
\be
A_T\Pra{\Pr_{L,T}} a^*,
\ee
where $\Pra{\Pr_{L,T}}$ stands for the convergence in probability with respect to $\Pr_{L,T}$ in the limit $T\ra\infty$. This notation is important -- it will be used later with other observables and path measures.

\subsection{Driven process}
\label{secdp1}

The definition of the driven process $\hX_t$ involves various large deviation elements related to $A_{T}$. The first is the \emph{scaled cumulant generating function} (SCGF) of $A_{T}$ defined as
\be
\Lambda_{k}=\lim_{T\ra\infty}\frac{1}{T}\ln \Ex_{\Pr_{L,T}}[e^{TkA_{T}}],
\label{eqscgfdef1}
\ee
where $k\in\reals$ and the expectation is taken with respect to the process $X_t$ with path measure $\Pr_{L,T}$. The second element that we need is the \emph{tilted generator}, given by
\be
\cL_{k}=\hF\cdot(\nabla+kg)+(\nabla+kg)\frac{D}{2}(\nabla+kg)+kf,
\label{eqtiltedgen1}
\ee
where $f$ and $g$ are the two functions entering in the definition of $A_{T}$ and $D$ is the diffusion matrix. This linear differential operator is essentially the generator of the evolution of the generating function of $A_T$, obtained by combining Girsanov's Theorem and the Feynman-Kac formula; see Sec.~3.1 and Appendix A.2 of \cite{chetrite2014}. It plays a central role in large deviation theory, as its dominant (Perron-Frobenius) eigenvalue coincides with the SCGF of $A_T$ if we assume that the spectrum of $\cL_k$ has a gap; see Sec.~3.2 of \cite{chetrite2014}. Under this assumption, we then have
\be
\cL_{k}r_{k}=\Lambda_{k}r_{k}.
\label{eqeig1}
\ee
where $r_k$ is the \emph{`right' eigenfunction} associated with the dominant eigenvalue and SCGF $\Lambda_k$. For the remaining, we also need the associated \emph{dual} or \emph{`left' eigenfunction} given by
\be
\cL^\dag_k l_k=\Lambda_k l_k.
\ee
These functions are normalized according to
\be
\int l_k(x)dx=1,\qquad\int l_k(x)r_k(x)dx=1.
\label{eq:norm}
\ee

With these elements, we define the \emph{driven process} $\hX_t$ as the Markov process with generator $L_{F_k}$ acting on functions $h$ according to
\be
L_{F_k}h = r_{k}^{-1} (\cL_{k} r_{k}h)-r_{k}^{-1} (\cL_{k} r_{k})h
= r_{k}^{-1} (\cL_{k} r_{k}h)-\Lambda_{k} h,
\label{eqgendoob1}
\ee
where $(\cL_{k} r_{k}h)$ means that $\cL_{k}$ is acting on the product $r_{k}h$. This transform of $\cL_{k}$ is a generalization of \emph{Doob's $h$-transform} \cite{doob1984,doob1957,rogers2000} which defines, as shown in \cite{chetrite2014}, a homogeneous Markov process with path measure 
\be
d\Pr_{L_{F_k},T}(\om)=\frac{r_{k}(X_{T}(\om))}{r_{k}(x_0)}e^{T[kA_T(\om)-\Lambda_{k}]}d\Pr_{L,T}(\om)
\label{eqpmdp1}
\ee
compared to the path measure of $X_{t}$.\footnote{We emphasize that this transform is a generalization of the Doob transform, since it relates $L_{F_k}$ and $L$ via the non-conservative Markov generator $\cL_k$. The resulting generator $L_{F_k}$ does however conserve probabilities.} The effect of this transform for diffusions \cite{chetrite2014} is to change only the drift $F$ of $X_{t}$ to the  \emph{driven drift}
\be
F_{k}=F+D(kg+\nabla\ln r_{k}),
\label{eqmoddrift1}
\ee
so that $\hX_t$ satisfies the SDE
\be
d\hX_t=F_{k}(\hX_t)dt+\sigma(\hX_t)\circ dW_{t}.
\ee
This explains the subscript $F_k$ in the generator of $\hX_t$.

The convergence of the driven process to the conditioned process $X_{t}|A_{T}=a$ follows from these results by assuming that (i) $A_T$ satisfies an LDP, (ii) the spectrum of $\cL_k$ is gapped, and (iii) the rate function $I(a)$ is convex.\footnote{There is a further technical assumption, namely, that the large deviations of $A_T$ do not arise as a boundary effect in time; see Secs.~5.2 and 5.3 of \cite{chetrite2014}.} Under these hypotheses, it can be shown that
\be
\lim_{T\ra\infty}\frac{1}{T}\ln \frac{d\Pr_{a,T}^{\micro}}{d\Pr_{L_{F_k},T}}(\om)=0
\label{eqproceq1}
\ee
for almost all path $\{X_{t}(\om)\}_{t=0}^{T}$ with respect to $\Pr_{a,T}$ or $\Pr_{L_{F_k},T}$ if we choose $k=I'(a)$. This limit establishes a form of process equivalence whereby 
\be
\Pr_{a,T}^{\micro}(d\om)\approx\Pr_{L_{F_k},T}(d\om)
\label{eqeq1}
\ee
with subexponential corrections in $T$, as in the expression (\ref{ldpdef1}) of the LDP, so that these two measures are equal on a logarithmic scale. In this sense, we say that the conditioned and driven processes become \emph{asymptotically equivalent} in the limit $T\ra\infty$ for $k$ such that $k=I'(a)$. This holds again if $I(a)$ is convex; if $I(a)$ is nonconvex, then there is no driven process that is equivalent to the conditioned process; see Sec.~5 of \cite{chetrite2014} for more detail.\footnote{We assume for simplicity that $I(a)$ is differentiable. If $I(a)$ is convex but not differentiable, then $k\in\partial I(a)$ where $\partial I(a)$ is the subdifferential of $I(a)$; see \cite{ellis2000} and Appendix 1 of \cite{touchette2015}.}

This process equivalence is a direct generalization of the equivalence of the microcanonical and canonical ensembles of equilibrium statistical mechanics \cite{touchette2015}. In fact, we prove the limit (\ref{eqproceq1}) in \cite{chetrite2014} in two steps using the following path generalization of the canonical ensemble:
\be
d\Pr_{k,T}^{\cano}(\om)=\frac{e^{TkA_{T}}}{\Ex_{\Pr_{L,T}}[e^{TkA_{T}}]}d\Pr_{L,T}(\om),
\label{eqcan1}
\ee
also known as the \emph{tilted path measure} or \emph{$s$-ensemble} \cite{lecomte2005,lecomte2007,garrahan2009,jack2010b}. First, we prove that the canonical path measure is described by a non-homogeneous (viz., time-dependent) Markov process which converges as $T\ra\infty$ to a homogeneous Markov measure with generator $L_{F_k}$. Second, we use known equivalence results for the canonical and microcanonical ensembles \cite{touchette2015} to establish a limit similar to (\ref{eqproceq1}) for the canonical and microcanonical path measures, which holds if $I(a)$ is convex and $k=I'(a)$. The limit (\ref{eqproceq1}) then follows using the chain rule for Radon-Nikodym derivatives:
\be
\frac{d\Pr_{a,T}^{\micro}}{d\Pr_{L_{F_k},T}}=\frac{d\Pr_{a,T}^{\micro}}{d\Pr_{k,T}^{\cano}} \frac{d\Pr_{k,T}^{\cano}}{d\Pr_{L_{F_k},T}}.
\ee

Physically, the equivalence of the conditioned and driven processes also means that these two different processes have the same typical states in the long-time limit. For equilibrium systems, it is known that observables in the microcanonical and canonical ensembles can have different fluctuations, but have the same equilibrium (viz.\ typical) values in the infinite-volume or thermodynamic limit when the microcanonical entropy is concave as a function of energy~\cite{ellis2000,touchette2015}. Similarly, it can be shown that time-integrated observables such as $A_T$ have in general different fluctuations with respect to the driven and conditioned processes, but have the same ergodic (viz.\ typical) values in the limit $T\ra\infty$ when $I(a)$ is convex \cite{chetrite2014}. Thus, although the conditioned process might not be Markov for $T<\infty$, it can be described in the ergodic limit by an effective Markov process~--~the driven process $\hX_t$~--~having the same typical values of observables. In particular, the `hard' constraint $A_T=a$ of the conditioned process is achieved in the driven process in a `soft' way via the limit
\be
A_T\Pra{\Pr_{L_{F_k},T}} a.
\ee 

This equivalence is proved in \cite{chetrite2014} for general observables, and holds in particular for two observables of mathematical and physical interest, namely, the \emph{empirical density}
\be
\rho_{T}(x)=\frac{1}{T}\int_{0}^{T}\delta (X_{t}-x)dt,
\ee
which represents the fraction of time spent at $x$, and the \emph{empirical current}
\be
J_{T}(x)=\frac{1}{T}\int_{0}^{T} \delta(X_{t}-x)\circ dX_{t},
\ee
which is a time-averaged local `velocity'. For ergodic processes, it is known that $\rho_T$ converges in probability to the stationary density, whereas $J_T$ converges in probability to the stationary Fokker-Planck current \cite{maes2008}. For the driven process, the stationary density  is \cite{chetrite2014}
\be
\rho^\inv_{F_k}(x)=r_k(x)l_k(x).
\ee
Consequently,
\be
\rho_T\Pra{\Pr_{L_{F_k},T}} \rho^\inv_{F_k},\qquad J_T\Pra{\Pr_{L_{F_k},T}} J_{F_k,\rho^\inv_{F_k}}
\ee 
for that process. The point of equivalence is that the same limits hold for the conditioned process when $I(a)$ is convex and $k$ and $a$ are related by $k=I'(a)$. In this case, the driven and conditioned processes have the same ergodic density and ergodic probability current. This is important for the following.

\section{Variational representations}
\label{secvarrep}

The driven process is intuitively the conditioning-free Markov process that realizes the conditioned process in the long-time limit. In this section, we propose three other ways to characterize this process using variational principles involving the SCGF of the conditioning observable, its rate function, and the relative entropy. These principles follow from general and simple large deviation arguments based on the contraction principle, which we first explain before deriving our main results. Some of these results were discussed in the literature, as mentioned in the introduction; the contribution of our work is to unify and to generalize them within the framework of large deviation conditioning and to show that their solutions give the driven process. 

\subsection{Main ideas}

It is known from large deviation theory that the SCGF and rate function of $A_T$ can be obtained from `higher' random variables $B_T$ satisfying two properties:

\begin{enumerate}
\item $B_{T}$ has an LDP with rate function $K(b)$. In our context, $B_T$ is an observable of the paths of $X_t$, as for $A_T$, so its LDP is also defined with respect to the path measure of $X_t$.
\item $A_{T}$ can be written as a function of $B_{T}$: that is, there exists a function $\tA$ such that 
\be
A_{T}(\om)=\tA(B_{T}(\om))
\ee
for all paths $\{X_{t}(\om)\}_{t=0}^{T}$.\footnote{The equality in this assumption can be weakened to $|A_{T}(\om)-\tA(B_{T}(\om))|=o(1)$ in $T$ as $T\ra\infty$; see \cite{ellis2000}.} In this case, we say that the observable $A_{T}$ admits a \emph{representation} or \emph{contraction} in terms of $B_{T}$.
\end{enumerate}

Under these assumptions, the following principles and equivalence result apply \cite{dembo1998,ellis2000,touchette2009}:
\begin{itemize}
\item \textbf{Contraction principle:}
\be
I(a)=\inf_{b: \tA(b)=a} K(b).
\label{eqcp2}
\ee
The solution $b^a$ of this constrained minimization\footnote{We assume for simplicity that the solution is unique; see \cite{ellis2000,touchette2015} for more detail about the case where more than one minimizers exist.} corresponds to the stationary value of $B_{T}$ on which $P\{B_T=b|A_T=a\}$ concentrates exponentially as $T\ra\infty$, so that
\be
B_T\Pra{\Pr_{a,T}^\micro}  b^a.
\ee 
This interpretation of $b^a$ follows by deriving from the two properties above an explicit rate function for $P\{B_T=b|A_T=a\}$; see \cite{ellis2000}, Sec.~5.3.2 of \cite{touchette2009}, and \cite{touchette2015}.

\item \textbf{Laplace principle:}
\be
\Lambda_{k}=\sup_{b}\{k \tA(b)-K(b)\}.
\label{eqcp1}
\ee
This is the Lagrange multiplier or dual version of the contraction principle. Its solution $b_k$, parameterized by $k$, corresponds to the typical value of $B_{T}$ under the canonical path measure (\ref{eqcan1}), which means
\be
B_T\Pra{\Pr_{k,T}^\cano} b_k;
\ee
see \cite{ellis2000} and Sec.~5.4 of \cite{touchette2009}. Since the canonical measure is equivalent to the path measure of the driven process, the limit above also holds for $\Pr_{L_{F_k},T}$; see Sec.~5.2 of \cite{chetrite2014} for more detail.

\item \textbf{Equivalence:} The contraction and Laplace principles have the same solutions for convex rate functions. More precisely, if $I(a)$ is convex at $a$, then $b^a=b_k$ for $k=I'(a)$. This follows from properties of Legendre-Fenchel transforms \cite{ellis2000,touchette2015} and is the basis of the equivalence between the conditioned and driven processes mentioned before \cite{chetrite2014}.
\end{itemize}

The idea for the rest of this section is to apply these results using special observables $B_T$ that can be related in a one-to-one way with path measures of Markov processes. In this case, the solutions of the variational problems (\ref{eqcp2}) and (\ref{eqcp1}) can be put in correspondence with a Markov process that turns out to be the driven process. Intuitively, this process can thus be interpreted as the unique Markov process that `minimizes' (\ref{eqcp2}) and `maximizes' (\ref{eqcp1}) via $B_{T}$. 

Contractions that can be used to derive these correspondences depend on the process and observable considered. For the class of Markov observables $A_T$ defined in (\ref{eqobs1}), there is a simple and general contraction involving the empirical density $\rho_T$ and empirical current $J_T$.\footnote{Another general contraction can be built from the so-called \emph{empirical process}, which is an abstract infinite-dimensional generalization of the empirical density $\rho_T$ defining in large deviation theory the \emph{level-3 LDP} \cite{ellis1985,deuschel1989,hollander2000}.\label{fn1}} Both are indeed known to satisfy, when considered jointly as $B_{T}=(\rho_{T},J_{T})$, an LDP with rate function 
\be
K(\rho,j)= 
\left\{
\begin{array}{lll}
\displaystyle\frac{1}{2}\int [j(x)-J_{F,\rho}(x)] (\rho D)^{-1}(x) [j(x)-J_{F,\rho}(x)] dx & & \text{if } \nabla\cdot j=0\\
\infty & & \text{otherwise.}
\end{array}
\right.
\label{eqratefct1}
\ee
Moreover, we have
\be
\tA(\rho_T,J_T)=\int f(x) \rho_{T}(x)dx+\int g(x)\cdot J_{T}(x) dx,
\label{eqrep1}
\ee
so that $A_T$ is a contraction of $B_{T}=(\rho_{T},J_{T})$.

This choice of $B_T$ defines in large deviation theory the \emph{level 2.5 of large deviations} or \emph{level-2.5 LDP} \cite{maes2008,maes2008a,chernyak2009,bertini2012,barato2015}. This is a natural large deviation level to consider for Markov diffusions, since the Radon-Nikodym derivative of any two homogeneous diffusions can be expressed exactly as a function of $\rho_{T}$ and $J_{T}$. This means essentially that these two random variables are sufficient to define a Markov process uniquely, which is the property needed to establish a relation between $B_{T}$ and the driven process. 

\subsection{SCGF}
\label{secsgf1}

We start by deriving variational representations for $\Lambda_k$ based on (\ref{eqcp1}). Using the rate function (\ref{eqratefct1}) and the representation function (\ref{eqrep1}) for the joint observable $(\rho_{T},J_{T})$, we first obtain
\be
\Lambda_{k}=\sup_{\rho,j} \left\{k\tA(\rho,j)-K(\rho,j)\right\}.
\label{eqvar1}
\ee
The maximization is performed over all normalized densities, $\int \rho(x)dx=1$, and currents $j$ satisfying the `sourceless' condition $\nabla\cdot j=0$ entering in the expression of $K(\rho,j)$. The link with the driven process is established by noting that the solution $(\rho^*,j^*)$ of this maximization is
\be
\rho^{*}=\rho^\inv_{F_k},\quad j^{*}= J_{F_{k},\rho^\inv_{F_k}},
\label{eqsol1}
\ee 
where $F_{k}$ is the modified drift of (\ref{eqmoddrift1}). This solution is derived explicitly in Appendix~\ref{secopt1} using Lagrange multipliers and can also be obtained in a simpler way by noticing, following our statement of the Laplace principle (\ref{eqcp1}), that $(\rho^*,j^*)$ corresponds to the ergodic value of $(\rho_T,J_T)$ in the driven process. Since we know from the previous section that this process is such that
\be
(\rho_T,J_T)\Pra{\Pr_{L_{F_k},T}} (\rho^\inv_{F_k},J_{F_k,\rho^\inv_{F_k}}),
\label{eqminsol1}
\ee
we must therefore have (\ref{eqsol1}).

Mathematically, the representation (\ref{eqvar1}) can be seen as a generalization of the spectral characterization of positive operators obtained by Donsker and Varadhan \cite{donsker1975b,donsker1975c,donsker1976b}, following Kac's derivation of a probabilistic formula for the smallest eigenvalue of Schrodinger operators \cite{kac1951}. Variants of the DV characterization were also obtained by Holland \cite{holland1978,holland1977}. As shown in Appendix~\ref{appdv}, the DV result is recovered for $g=0$ by direct minimization over $j$, leaving in (\ref{eqvar1}) only the empirical density whose large deviations are referred to as the level-2 of large deviations. Moreover, as mentioned in \cite{donsker1975b} and shown here in Appendix~\ref{appdv}, the resulting variational characterization of $\Lambda_{k}$ further reduces to the Rayleigh-Ritz principle \cite{courant2004}, commonly used in quantum mechanics, in the case where $L=L^\dag$ with respect to the Lebesgue measure.

The relation between the variational principle (\ref{eqvar1}) and the driven process can be made more explicit using the fact that $\rho$ and $j$ uniquely determines the drift. To this end, let us rewrite the maximization in (\ref{eqvar1}) by expressing the current fluctuation $j$ as $j=J_{u,\rho}$, where $J_{u,\rho}$ is the stationary current (\ref{eqcurr1}) associated with a `free' drift $F=u$ and diffusion matrix $D$. The constraint $\nabla\cdot j=0$ implies that $\rho$ is the invariant density $\rho^\inv_u$ of a process with drift $u$ and diffusion $D$. Changing the variables $(\rho,j)\ra(\rho,u)$ then leads to
\be
\Lambda_{k}=\sup_{u}\left\{ k \tA(\rho_u^\inv,J_{u,\rho_u^\inv})-K(\rho_u^\inv,J_{u,\rho_u^\inv})\right\},
\label{eqvar2}
\ee  
where
\be
K(\rho_u^\inv,J_{u,\rho_u^\inv})=\frac{1}{2}\int [u(x)-F(x)]D^{-1}(x)[u(x)-F(x)] \rho_u^\inv(x)\, dx
\label{eqrfu1}
\ee
is the level-2.5 rate function (\ref{eqratefct1}) expressed via (\ref{eqcurr1}) in terms of the drift $u$. Moreover, because of (\ref{eqminsol1}) and the change of variables to $u$, the maximizer is now $u^*=F_k$.\footnote{Unlike $(\rho^*,j^*)$, $u^*$ cannot be interpreted as a `most probable drift' -- it is simply the drift that makes the solution $(\rho^*,j^*)$ of (\ref{eqvar1}) typical.}

This representation of the SCGF is a level-2.5 generalization of a control result of Fleming \cite{fleming1992} discussed in more detail in Sec.~\ref{seccont}. It also generalizes previous results obtained for 1-$d$ diffusions, which are characterized by a constant current because of the sourceless condition $\nabla\cdot j=0$. In particular, (\ref{eqvar2}) generalizes Eq.~(4.9) of Sughiyama and Ohzeki \cite{sughiyama2013} who consider the special observable $A_T=X_T/T$, obtained here with $f=0$ and $g=1$. Additionally, it recovers Eq.~(12) of Nemoto and Sasa \cite{nemoto2011b} (see also \cite{nemoto2011}) who consider the same observable for an overdamped Langevin equation on a ring of circumference $L$, driven by a constant drive $f$, a periodic force derived from a potential $U$, and a heat bath (Gaussian white noise) with inverse temperature $\beta$. Their main result for this model, written in our notations $D=2/(\beta\gamma)$ and $F=(f-U')/\gamma$, is
\be
\Lambda_k=-\frac{\beta}{4} \inf_{u} \left\{ \int_0^L dx \left(\rho^\inv_{u/\gamma+F}(x)\frac{u(x)^2}{\gamma}-2 J_{u/\gamma+F,\rho^\inv_{u/\gamma+F}} u(x)\right)\right\}
\label{eqsasa1}
\ee
with the minimization constraint
\be
\int_0^L u(x)dx=\frac{2kL}{\beta}.
\label{eqcons1}
\ee

This follows from our result (\ref{eqvar2}) by setting $J_{u,\rho^\inv_u}$ to a constant, which yields
\be
\Lambda_k=-\frac{\beta}{4}\inf_u \left\{ \int_0^Ldx\, \rho_{u/\gamma+F}^\inv(x) \frac{u(x)^2}{\gamma}-\frac{4kL}{\beta} J_{u/\gamma+F,\rho^\inv_{u/\gamma+F}}\right\}
\label{eqoursasa1}
\ee
after the change of variables $u\ra u/\gamma+F$, and by noting that the minimizer
\be
u^*= \gamma (F_k-F)=\frac{2}{\beta}[k+(\ln r_k)']
\ee  
satisfies the integral constraint (\ref{eqcons1}) for all parameters.  Therefore, we do not change the solution of the minimization in (\ref{eqoursasa1}) by considering drifts $u$ for which (\ref{eqcons1}) holds. Using this relation in (\ref{eqoursasa1})  to replace part of the factor in front of the current by an integral and inserting the constant current inside that integral then yields (\ref{eqsasa1}). 

This derivation shows that the constraint (\ref{eqcons1}) is not fundamental: the variational problem to solve for general processes and observables involves, as shown in (\ref{eqvar2}), only a maximization over $u$ which can be interpreted as a control drift, as explained in the next section. For another study of the ring model focusing on current conditioning, see \cite{chetrite2013}. 

\subsection{Rate function}

Representations of the driven process can be obtained for the rate function $I(a)$ in a dual way from the variational representation (\ref{eqcp2}). Using the explicit rate function (\ref{eqratefct1}) and the contraction (\ref{eqrep1}) for $(\rho_T,J_T)$ in (\ref{eqcp2}) yields
\be
I(a)=\inf_{(\rho,j): \tA(\rho,j)=a} K(\rho,j).
\label{eqvarfct1}
\ee
If $I(a)$ is convex, the equivalence between the conditioned and driven processes implies that the solution of the minimization above is the same as the solution of the maximization (\ref{eqvar1}) giving $\Lambda_k$: that is, $\rho^*=\rho_{F_k}^\inv$ and $j^*=J_{F_k,\rho^\inv_{F_k}}$, with $k$ chosen so that
\be
\tA(\rho_{F_k}^\inv,J_{F_k,\rho^\inv_{F_k}})=a
\ee 
or equivalently $k=I'(a)$. This recovers the equivalence result mentioned at the end of Sec.~\ref{secnot} about the typical values $\rho_T$ and $J_T$ being the same in the conditioned process with $A_T=a$ and the driven process. A different proof of this result, which mimics the proof of Appendix~\ref{secopt1}, follows by solving (\ref{eqvarfct1}) with a Lagrange multiplier and by relating this multiplier to the constraint $\tA(\rho,j)=a$. 

Similarly to (\ref{eqvar2}), we can re-express the minimization in (\ref{eqvarfct1}) for a fixed diffusion matrix $D$ by the drift $u$ to obtain
\be
I(a)=\inf_{u:\tA(\rho_u,J_{u,\rho_u^\inv})=a} K(\rho_u^\inv,J_{u,\rho_u^\inv})
\label{eqvarfct2}
\ee
with $K(\rho_u^\inv,J_{u,\rho_u^\inv})$ shown in (\ref{eqrfu1}). As before, the minimizer is $u^*=F_k$ with $k=I'(a)$ if the rate function $I(a)$ is convex at $a$. The interpretation of this result is that the driven process is the controlled Markov process which maximizes the probability of $(\rho_T,J_T)$ under the constraint $A_T=a$. The jump process versions of these results are discussed in Appendix~\ref{appjp}.

\subsection{Relative entropy}
\label{secre1}

The variational representations derived before can be re-expressed at the more fundamental level of path measures using the notion of relative entropy. We recall that, given two probability measures $P$ and $Q$, the \emph{relative entropy} of $P$ with respect to $Q$ is defined as
\be
S(P||Q)=\int dP(\om) \ln \frac{dP}{dQ}(\om)=\Ex_P\left[\ln \frac{dP}{dQ}\right],
\ee
where $dP/dQ$ denotes the \emph{Radon-Nikodym derivative} of $P$ with respect to $Q$. This quantity is such $S(P||Q)\geq0$ with equality if and only if $P=Q$ almost everywhere. As a result, it is often used as a distance between probability measures, called the \emph{Kullback-Leibler distance}, even though it is not symmetric and does not satisfy the triangle identity \cite{cover1991}.

We state next the variational representations of $\Lambda_k$ and $I(a)$ obtained with the relative entropy and then discuss their meaning, proofs, and equivalence with the previous representations.

The first representation for the SCGF is
\be
\Lambda_k=\lim_{T\ra\infty}\sup_u \left\{k\, \Ex_{\Pr_{L_u,T}}[A_T]-\frac{1}{T}S(\Pr_{L_u,T}||\Pr_{L,T})\right\}
\label{eqrerep1}
\ee
and involves the relative entropy between the path measure $\Pr_{L_u,T}$ of a diffusion with drift $u$ and the path measure $\Pr_{L,T}$ of the original diffusion with drift $F$.\footnote{The diffusion matrix is $D$ for both processes, so they are equivalent in the sense of absolute continuity.} As in (\ref{eqvar2}), the solution of the maximization is $u^*=F_k$, the drift of the driven process.

The second representation is the dual of the formula above:
\be
I(a)=\lim_{T\ra\infty}\inf_{u: \Ex_{\Pr_{L_u,T}}[A_T]=a} \frac{1}{T}S(\Pr_{L_u,T}||\Pr_{L,T}).
\label{eqrerep2}
\ee
The `optimal' drift that solves this constrained minimization is $u^*=F_{k}$ as in (\ref{eqvarfct2}) with $k=I'(a)$ for $I(a)$ convex. This result is interesting physically -- it shows that the driven process is the homogeneous Markov process \emph{closest} to $X_t$, in the sense of relative entropy, that satisfies the constraint $\Ex_{\Pr_{L_u,T}}[A_T]=a$ or, equivalently, that makes $A_T=a$ typical in the long-time limit. 

This way of expressing a rate function via a change of measure that transforms the fluctuation $A_T=a$ for $X_t$ into a typical event for a modified process is very common in large deviation theory: it is basis of the so-called \emph{tilting method} (see Appendix C.2 of \cite{touchette2009} and Sec.~9.6 of \cite{varadhan2003}) and the weak-convergence approach to large deviations proposed by Dupuis and Ellis \cite{dupuis1997} for discrete-time processes. Conceptually, the relative entropy representations (\ref{eqrerep1}) and (\ref{eqrerep2}) can also be seen as a contraction of the level 3 of large deviations mentioned in the footnote~\ref{fn1}. For an explanation of this level in the simple case of independent random variables, see Sec.~II.5 of \cite{hollander2000}; for continuous-time processes, see Exercise 4.4.41 of \cite{deuschel1989}.

Physically, the path representation (\ref{eqrerep2}) is also interesting because it provides a probabilistic interpretation of the maximum entropy or maximum caliber approach to nonequilibrium systems mentioned in the introduction. Indeed, the process obtained from (\ref{eqrerep2}) is not only the process that minimizes the path relative entropy subject to a constraint -- it is the process that gives the rate function of $A_T$ as a result of this constrained minimization, as well as the process that one obtains in the long-time limit by conditioning, in the probabilistic sense, the original process $X_t$ on the constraint $A_T=a$. Some of these links were noted by Evans~\cite{evans2005a} and by Monthus \cite{monthus2011} for specific observables of jump processes and Markov chains; see also Appendices~\ref{appjp} and \ref{appmc} of this work.

We give next two proofs of the relative entropy representations (\ref{eqrerep1}) and (\ref{eqrerep2}): a simple proof based on the positivity of the relative entropy, and a slightly more involved proof that has the advantage of clarifying the relationship between the generalized Doob transform, the canonical path measure, and the maximum caliber method. A third and more direct proof, which uses the level 2.5 of large deviations, is given in Appendix~\ref{appreproof1}.

The first proof proceeds from the following limit:
\be
\lim_{T\ra\infty}\inf_u \frac{1}{T} S(\Pr_{L_u,T}|| \Pr_{L_{F_k},T})=0,
\label{eqresimple1}
\ee
which holds trivially because of the positivity of the relative entropy and the fact that the relative entropy is zero for $u=F_k$. Inserting the expression (\ref{eqpmdp1}) of $\Pr_{L_{F_k},T}$ in this limit and neglecting the end-point terms of this path measure involving $r_k$, which are finite by assumption, we directly obtain the relative entropy principle (\ref{eqrerep1}) and its constrained version (\ref{eqrerep2}) by duality. 

This is by far the simplest proof that we have of the relative entropy principles and, in fact, of all the variational representations derived in the previous sections, since these are equivalent to the relative entropy representations, as shown in Appendix~\ref{appreproof1}. The simple limit (\ref{eqresimple1}) thus provides a simple and powerful way to obtain large deviation functions at the path level, and can also be used as an alternative method for proving the control results of Sec.~\ref{seccont}.

The second proof of (\ref{eqrerep1}) and (\ref{eqrerep2}) proceeds differently. It uses the well-known fact that the canonical path measure $\Pr_{k,T}^\cano$ defined in (\ref{eqcan1}) is the unique solution for all $T<\infty$ of the following variational problem:
\be
\frac{1}{T}\ln \Ex_{\Pr_{L,T}}[e^{kTA_T}] = \sup_{\Qr_T}\left\{k\, \Ex_{\Qr_{T}}[A_T]-\frac{1}{T}S(\Qr_T|| \Pr_{L,T})\right\},
\label{eqscgfvar1}
\ee
where $\Qr_T$ is any path measure (not necessarily Markovian) over the time interval $[0,T]$ \cite{pra1996b}. This problem is a variant of the Gibbs or Kullback inequality. Assuming that the limit (\ref{eqscgfdef1}) defining the SCGF exists, we then obtain \cite{sughiyama2013}
\begin{eqnarray}
\Lambda_k &=& \lim_{T\ra\infty} \sup_{\Qr_T}\left\{k\, \Ex_{\Qr_{T}}[A_T]-\frac{1}{T}S(\Qr_T|| \Pr_{L,T})\right\}\nonumber\\
& = & \lim_{T\ra\infty}\left\{k\, \Ex_{\Pr_{k,T}^\cano}[A_T]-\frac{1}{T}S(\Pr_{k,T}^\cano|| \Pr_{L,T})\right\}.
\label{eqscgfvar2}
\end{eqnarray}

The driven process comes into this by noting that the canonical path measures $\Pr_{k,T}^\cano$ defines a \emph{non-homogeneous} Markov process with time-dependent generator, which becomes asymptotically equivalent with the \emph{time-homogeneous} driven process $\hX_t$ with drift $F_k$ in the long-time limit, so that
\be
\lim_{T\ra\infty}\frac{1}{T}S(\Pr_{L_{F_k},T}||\Pr_{k,T}^\cano)=0.
\label{eqproceqre1}
\ee
This process equivalence is proved in \cite{chetrite2014} and implies that the canonical path measure can be replaced in (\ref{eqscgfvar2}) by the driven process: both have the same ergodic states and the same asymptotic relative entropy relative to $X_t$, so that (\ref{eqscgfvar2}) is equivalent to (\ref{eqrerep1}) with $u^*=F_k$. This is also evident by comparing (\ref{eqproceqre1}) with (\ref{eqresimple1}). In the end, the driven process $\hX_t$ can thus be characterized as the homogeneous Markov process closest, in the sense of relative entropy, to the non-homogeneous canonical path measure. Because of the equivalence between the canonical and microcanonical path measures expressed in (\ref{eqproceq1}), it is also the homogeneous Markov process closest, in the sense of relative entropy, to the conditioned process $X_t|A_T=a$.

\subsection{Spectral eigenfunctions}

We close this section by presenting three other variational representations that characterize the spectral elements $r_k$ and $l_k$ rather than the driven process itself. The first is obtained by applying a change of variables to the variational principle (\ref{eqvar2}) which yields, as shown in Appendix~\ref{appserep1},
\be
\Lambda_k=\sup_{h>0}\int dx\, \rho^\inv_{F+D(kg+\nabla\ln h)}(x)\, h^{-1}(x)(\cL_k h)(x).
\label{eqrepse1}
\ee
The maximizer is $h^*=r_k$. This is a generalization of a result derived by Nemoto and Sasa for jump processes; see Appendix G of \cite{nemoto2011}. A more direct proof of this result follows by solving the maximization using functional derivatives and by finding that $h^*=r_k$ is the global maximum.

The second representation, obtained in the special case $g=0$, is
\be
\Lambda_k =\statpt_{l,r:\int l(x) r(x)\, dx=1} \int l(x) (\cL_k r)(x)\, dx,
\label{eqrepse2}
\ee
where $\statpt$ stands for the stationary point(s) of the expression on the right side, which are explicitly $l^*=l_k$ and $r^*=r_k$. The dual of this result gives $I(a)$ as
\be
I(a)=-\statpt_{l,r} \int l(x) (Lr)(x)\, dx
\label{eqrepse3}
\ee
subject to the two constraints
\be
\int l(x) r(x)\, dx=1\qquad\text{and}\qquad\int l(x) r(x) f(x)\, dx = a.
\label{eqrepse3cons}
\ee
This holds again for $g=0$ and is solved as before for $l^*=l_k$ and $r^*=r_k$.

The variational principles (\ref{eqrepse2}) and (\ref{eqrepse3}) are derived in Appendix~\ref{appserep2} from our previous representations involving $(\rho,j)$. The last one was previously derived using different methods by Eyink \cite{eyink1996a} (see also Symanzik \cite{symanzik1970}), who refers to it as the \emph{action principle} generalizing the Rayleigh-Ritz principle. We reproduce it here for completeness. Applications of this result have been studied in the context of turbulence \cite{eyink1998} and a simple Kramers model of nonequilibrium systems \cite{alexander1997}.

The representation (\ref{eqrepse2}) is obviously weaker than (\ref{eqrepse1}), since it applies only for $g=0$ and involves an optimization on two functions, compared to one in (\ref{eqrepse1}), which does not necessarily yield a global maximum. From (\ref{eqrepse1}), it is tempting to think that additional conditions (e.g., convexity of the rate function) might strengthen (\ref{eqrepse2}) and (\ref{eqrepse3}) to yield a global maximum or minimum rather than a stationary point. We have not been able to obtain any result in that direction. We also do not know whether the two representations involving stationary points above can be generalized to current-type observables with $g\neq 0$.

\section{Optimal control representations}
\label{seccont}

The variational principles derived in the previous section, especially those expressed in terms of the drift $u$, suggest that the driven process is a control process optimizing functionals related to the large deviations of $A_T$. We formalize this idea in this section by defining a controlled process explicitly and by showing that the functionals optimized have the form of `empirical' or `running' costs accumulated over the time interval or \emph{horizon} $[0,T]$. Moreover, we show that the so-called \emph{value function}, corresponding to the optimal control cost, satisfies a HJB equation and yields the optimal control drift, which converges in the ergodic limit to the driven drift. Relations between these results, those of Fleming and collaborators \cite{fleming1971,fleming1978,fleming1983,sheu1984,fleming1985b,fleming1987,sheu1991,fleming1992b,fleming1992,fleming1995,fleming1997}, and the more recent work of Bierkens, Chernyak, Chertkov and Kappen~\cite{chernyak2014,bierkens2013} are discussed at the end of the section. 

\subsection{Controlled process}

We consider the problem of maximizing the cost functional,
\be
C_t^T[X^u,u] = k\int_t^T \left[f(X^u_s)ds +g(X^u_s)\circ dX_s^u\right]
 - \frac{1}{2}\int_t^T (u_s-F) D^{-1} (u_s-F)(X^u_s)\ ds
\label{eqcost1}
\ee
for the controlled SDE 
\be
dX_t^u=u_t dt+\sigma(X_t^u)\circ dW_t,\qquad X_t\in\reals^d,
\label{eqdiff1}
\ee
driven by the \emph{control drift} $u_t$. The cost function $C_t^T$ comes from the variational principles of the previous section, in particular from~(\ref{eqvar2}): the first integral measures with the Lagrange multiplier $k$ the cost of reaching the constraint $A_T=a$, expressed now over a time-interval $[t,T]$ rather than $[0,T]$, while the second integral measures according to (\ref{eqrelurf1}) the cost of the control as the relative entropy between the controlled process $X_t^u$ with drift $u$ and the original uncontrolled process $X_t$ with drift $F$. Importantly for the theory, this cost is additive in time and quadratic in the control drift $u_t$. 

Stochastic control theory \cite{bellman1967,fleming1975,stengel1994,fleming2006,schulz2006} is concerned with determining the optimal control strategy $\{u_s^*\}_{s=t}^T$ that maximizes the expected cost starting with the initial condition $X_t=x$. The \emph{optimal} expected cost function is called the \emph{value function} and is denoted here by $\Lambda_t^T(x,k)$. Thus,
\be
u^*=\arg\sup_u \Ex_{\Pr_{L_u},T}[ C_t^T]
\label{eqoptcont1}
\ee
and
\be
\Lambda_t^T(x,k)=\sup_u \Ex_{\Pr_{L_u,T}}[ C_t^T],
\label{eqoptcost1}
\ee
where the expectation is with respect to the control process $X^u_t$ started at $X_t^u=x$.

\subsection{Optimal controller}

The solutions of (\ref{eqoptcont1}) and (\ref{eqoptcost1}) are obtained from standard results of stochastic control theory adapted in Appendix~\ref{apphjb} to control costs containing a displacement or current cost with $g\neq 0$. The control problem involving these costs satisfies the dynamic programming principle of Bellman which leads to the following stochastic \emph{Hamilton-Jacobi-Bellman} (HJB) equation:
\be
-\p_t \Lambda_t^T=\sup_u\left\{ kf+kg\cdot \hat u-\frac{1}{2}(u-F)D^{-1}(u-F)+\frac{k}{2}\nabla \cdot (D g)+L_u \Lambda_t^T\right\}
\label{eqhjb1}
\ee
with final condition $\Lambda_T^T=0$. This equation is derived in Appendix~\ref{apphjb}; it involves the Markov generator $L_u$ of the controlled diffusion $X_t^u$ defined in (\ref{eqdiff1}) and reduces for $g=0$ to the standard stochastic HJB equation. The optimal control law (\ref{eqoptcont1}) is the maximizer of this equation and is given, after inserting the expression of the generator $L_u$ and performing the maximization, by 
\be
u^*_t=F+D(kg+\nabla\Lambda_t^T).
\label{eqver1}
\ee
This is a non-homogeneous optimal control law that depends explicitly, as is normal in control theory, on the initial and final times of the control horizon $[t,T]$. Putting $g=0$ yields the standard solution $u^*_t=F+D\nabla\Lambda_t^T$ \cite{fleming2006}.

To obtain the long-time limit of these results, we consider the exponential cost $G_t^T(x,k)=e^{\Lambda_t^T(x,k)}$.
We show in Appendix \ref{appcontrep} that the new function $G_t^T(x,k)$ obtained from this Hopf-Cole transform satisfies the \emph{backward Feynman-Kac equation}
\be
(\p_t +\cL_k) G_t^T=0
\label{eqhjb2}
\ee 
with $G_T^T=1$ and $\cL_k$, the tilted generator defined in (\ref{eqtiltedgen1}). The  solution of this equation is obtained formally by integrating the semi-group generated by $\cL_k$ at the final position $X_T=y$:
\be
G_t^T(x,k)=\int e^{(T-t)\cL_k}(x,y)\, dy.
\ee
Using the assumption that the spectrum of $\cL_k$ has a gap, denoted by $\Delta_k$, it can then be shown (see \cite{chetrite2014} for details) that $G_t^T(x,k)$ behaves exponentially for large $T-t$ according to
\be
G_t^T(x,k)=r_k(x)e^{(T-t)\Lambda_k}[1+O(e^{-(T-t)\Delta_k})].
\ee
 Consequently,
\be
\lim_{T\ra\infty} \frac{\Lambda_t^T(x,k)}{T}=\lim_{T\ra\infty}\frac{1}{T}\ln G_t^T(x,k)=\Lambda_k
\label{eqcostlim1}
\ee
and
\be
\lim_{T\ra\infty} \nabla \Lambda_t^T(x,k)=\lim_{T\ra\infty}\frac{\nabla G_t^T(x,k)}{G_t^T(x,k)}=\frac{\nabla r_k(x)}{r_k(x)}=\nabla \ln r_k(x),
\ee
so that
\be
\lim_{T\ra\infty} u^*_t=F+D(kg+\nabla\ln r_k)=F_k
\ee
by (\ref{eqver1}) and (\ref{eqmoddrift1}). This shows that the driven process is the optimal control process maximizing the expectation of the cost $C_t^T$ as $T\ra\infty$. To be more precise, it is the optimal process that maximizes by (\ref{eqcostlim1}) the \emph{mean} expected cost, which converges to $\Lambda_k$ in the ergodic limit.

We can strengthen this result slightly using convergence in probability instead of convergence in mean by considering the \textit{time-rescaled} cost $\Lambda_t^T/T$, which converges according to (\ref{eqcostlim1}) to the SCGF and dominant eigenvalue $\Lambda_k$. Thus,
\be
\Lambda_k=\lim_{T\ra\infty} \sup_{u} \frac{1}{T}\Ex_{\Pr_{L_u,T}}[C_0^T] =\lim_{T\ra\infty} \sup_u \Ex_{\Pr_{L_u,T}}[ kA_T-K_T\big],
\label{eqcontvar2}
\ee
where $A_T$ is our usual observable evaluated for $X_t^u$ and
\be
K_T=\frac{1}{2T}\int_0^T (u-F)D^{-1}(u-F)(X_t^u)\, dt 
\label{eqktd1}
\ee
is an `empirical' or `sample mean' version of the level-2.5 rate function shown in (\ref{eqrfu1}). Noting that
\be
K_T\Pra{\Pr_{L_u,T}}K(\rho_u^\inv,J_{u,\rho_u^\inv})\quad\text{and}\quad A_T\Pra{\Pr_{L_u,T}}\tA(\rho_u^\inv,J_{u,\rho_u^\inv}),
\ee
we can then rewrite (\ref{eqcontvar2}) as 
\be
\Lambda_k= \lim_{T\ra\infty}\sup_u \{kA_T-K_T\},
\label{eqcontd1}
\ee
which is now understood as a limit in probability with respect to the law of $X_t^u$. The optimal controller solving this maximization is $u^*=F_k$.

\subsection{Comparison with previous works}
\label{seccomp}

The derivation above follows essentially the theory of Fleming and collaborators \cite{fleming1971,fleming1978,fleming1983,sheu1984,fleming1985b,fleming1987,sheu1991,fleming1992b,fleming1992,fleming1995,fleming1997} relating linear PDEs to control problems. The main difference is our consideration of a new $g$ term in the cost related to observables $A_T$ that depend on the displacements or increments of $X_t$ in addition to the state $X_t$ itself. For $g=0$, it can be checked that the expected cost appearing in (\ref{eqcontvar2}) with (\ref{eqktd1}) is equivalent to the cost (3.10) obtained by Fleming in \cite{fleming1992}.

Conceptually, our approach is also a reversal of Fleming's, in that we transform the nonlinear HJB PDE (\ref{eqhjb1}) into the linear Feynman-Kac PDE (\ref{eqhjb2}). Moreover, we provide, as mentioned in the introduction, a new probabilistic interpretation of the optimal control process arising in Fleming's theory as the process optimizing the cost $C_t^T$. In the ergodic limit, this process is the driven process $\hX_t$ and, by equivalence, the conditioned process $X_t|A_T=a$.

This not only complements Fleming's theory, but provides, as also mentioned in the introduction, a different approach to the recent work of Bierkens, Chernyak, Chertkov, and Kappen~\cite{chernyak2014,bierkens2013}, who generalize optimal control theory to current-type costs by working directly with ergodic controls for the cost (\ref{eqcost1}). More precisely, they consider a variational principle similar to (\ref{eqvar2}) involving a control drift $u$ (Problem 2.3 in \cite{bierkens2013}), which they transform to a variational principle of the type (\ref{eqvar1}) for $\rho$ and $j$ (Problem 3.13 in \cite{bierkens2013}). From the latter principle, they then deduce an equation for the maximizers $\rho^*$ and $j^*$, which they call the HJB equation, and apply a Hopf-Cole transform to obtain a linear equation (in Theorem 5.10 of \cite{bierkens2013}) which is essentially our  equation (\ref{eqeig1}) defining $r_k$ and $\Lambda_k$. In doing so, they observe that they generalize the DV characterization of dominant eigenvalues of positive operators, but they do not identify the optimal control as the driven or conditioned process.

There are two other connections worth mentioning. The first is with the weak-convergence approach of Dupuis and Ellis \cite{dupuis1997}, alluded to in Sec.~\ref{secre1}, which is very close conceptually to our derivation of the SCGF based on the canonical path measure and the variational problem (\ref{eqscgfvar1}). The second connection is with Bierkens and Kappen \cite{bierkens2014}, who solve the optimization problem
\be
J^*=\inf_{\Qr} \{ \Ex_{\Qr} [C]+S(\Qr||\Pr)\}
\label{eqbkopt1}
\ee
for general probability measures $\Qr$ and $\Pr$, including path measures of Markov processes, and notice that the solution is a canonical measure (see  \cite{pra1996b} for related results). This is consistent with both the relative entropy representation (\ref{eqrerep1}) derived in the previous section, which has the form (\ref{eqbkopt1}) and which yields the driven process as the ergodic limit of the canonical path measure, and the control results of this section, which express this representation via a drift-controlled process. It seems in fact that some of the optimal control processes obtained in \cite{bierkens2014} can be expressed as a generalized Doob transform similar to (\ref{eqgendoob1}).

\section{Conclusions}
\label{seccon}

\begin{figure}
\centering
\resizebox{0.82\textwidth}{!}{\includegraphics{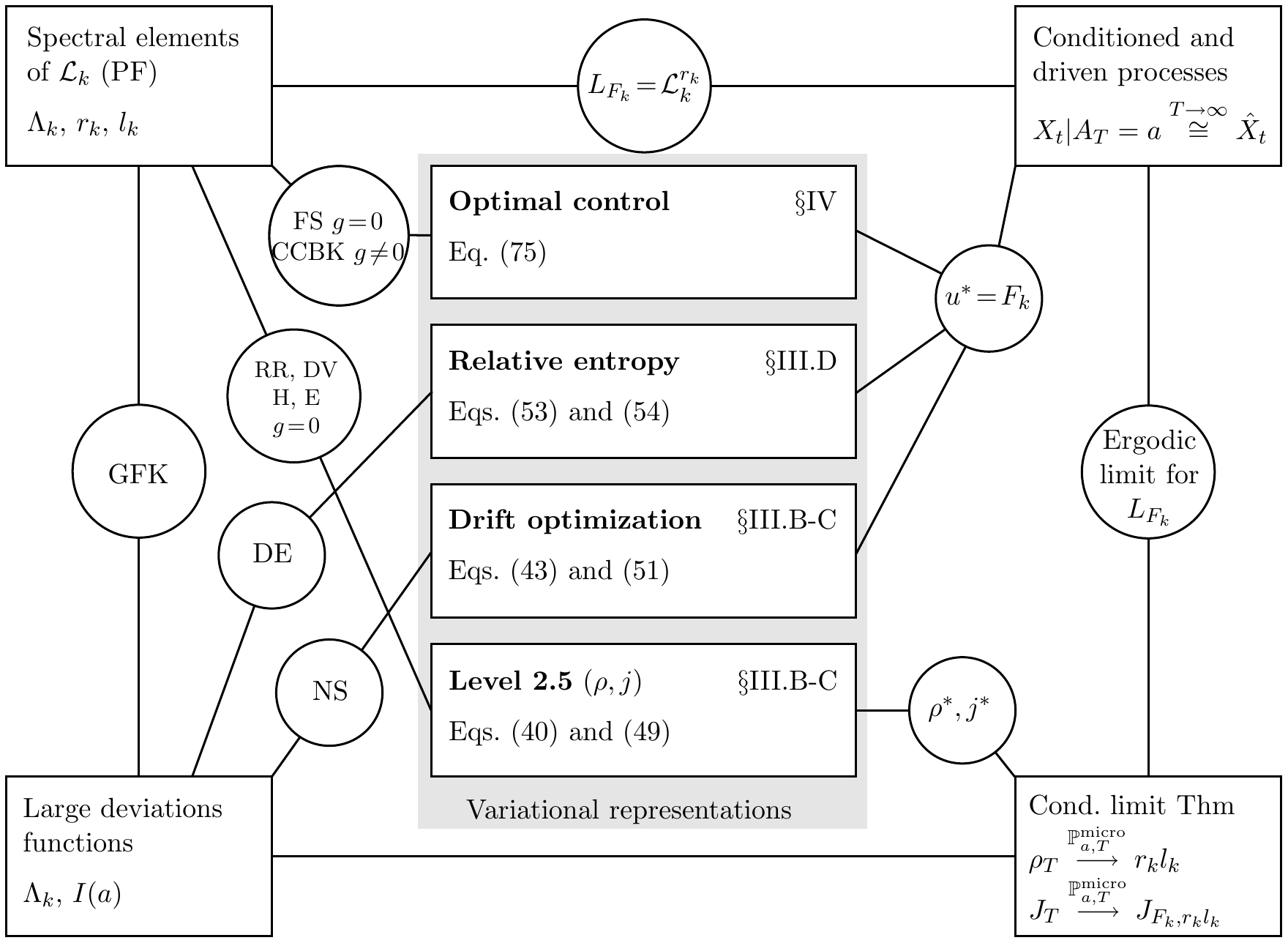}}
\caption{Summary diagram of our results with important links. Acronyms: PF~=~Perron-Frobenius, FS~=~Fleming-Sheu, CCBK = Chernyak, Chertkov, Bierkens, Kappen, RR~=~Rayleigh-Ritz, DV~=~Donsker-Varadhan, H~=~Holland, E~=~Eyink, GFK~=~Girsanov-Feynman-Kac, DE~=~Dupuis-Ellis, NS~=~Nemoto-Sasa; see the text for more links and references.}
\label{figtable1}
\end{figure}

We give in Fig.~\ref{figtable1} a summary diagram containing the main results of this paper in the center and various links between these results, the driven process, and the elements (spectral and large deviation) used to construct that process. In order not to overfill the diagram, we include on the left-hand side of the diagram links with previous works that directly motivated this paper; more links are explained in the text. Our own contributions, which occupy the center and right-hand side of the diagram, are
\begin{itemize}
\addtolength{\itemsep}{-5pt}
\item to derive general variational principles or representations for $\Lambda_k$ and $I(a)$, based on the level 2.5 of large deviations, generalizing previous representations obtained by Nemoto and Sasa \cite{nemoto2011,nemoto2011b,nemoto2014} and by Eyink \cite{eyink1996a,alexander1997,eyink1998} among others;
\item to explicitly link these variational principles to the control approach to large deviations developed by Fleming \cite{fleming1971,fleming1978,fleming1983,sheu1984,fleming1985b,fleming1987,sheu1991,fleming1992b,fleming1992,fleming1995,fleming1997}; 
\item to show that the solution of these variational principles and control problems is the driven process and, when equivalence holds, the conditioned process.
\end{itemize}

We have focused in the previous sections on deriving these principles and explaining how they follow from the contraction principle and Laplace principle of large deviation theory. In the remaining, we discuss three important applications related to the physics of large deviations in nonequilibrium systems and approximations (analytical or numerical) of large deviation functions. The discussion is meant to be brief; our goal is to give a few remarks pointing to how the variational representations derived here can be used to study nonequilibrium systems and to compute large deviation functions describing their fluctuations. We plan to develop each of these remarks, especially those related to numerical methods for estimating large deviation functions, more extensively in a series of future publications. The full implementation of these methods, and their comparison with other methods such as cloning \cite{lecomte2007a,tailleur2009,giardina2011}, represent an important and challenging problem for nonequilibrium statistical physics and large deviation theory as a whole.

\subsection{Physics of nonequilibrium fluctuations}
 
We have already mentioned in the introduction that our theory of conditioned and driven processes can be seen as generalization of the FWG theory \cite{freidlin1984,graham1989,luchinsky1998} when applied to time-integrated observables. The starting point of both theories is the same: to describe how fluctuations `arise’ or `are created’ spontaneously from noise. Moreover, both refer to a conditioning problem. In the FWG theory, this conditioning selects a single path, called the \emph{fluctuation path} or \emph{instanton}, because of the large deviation form of the path measure in the low-noise limit \cite{freidlin1984}. By contrast, in our theory the conditioning does not select a single path in general, but a whole set of paths or \emph{process} identified as the driven process $\hX_t$ in the ergodic limit.

We have seen another interesting way to characterize the driven process, namely, as a process that transforms a fluctuation into a typical state. In general, many different stochastic processes can be used to achieve this process transformation or path \emph{reweighting}. The driven process is special among reweightings in that it is the unique homogeneous Markov process \emph{closest to the original process $X_t$, with respect to the `distance measure' defined by the relative entropy}, which transforms $X_t$ to make the fluctuation $A_T=a$ typical. This provides, as mentioned, a first-principle justification of the maximum entropy approach to nonequilibrium systems, which was proposed as an \emph{ad hoc} generalization of the notion of statistical ensembles to these systems \cite{jaynes1980,stock2008,presse2013}.

Many open problems remain about the application of the driven process to study nonequilibrium systems beyond calculating large deviation functions. In particular, 
\begin{itemize}
\addtolength{\itemsep}{-5pt}
\item Can nonequilibrium systems be represented as a conditioning of equilibrium systems? Consider, to be more precise, an extended many-particle  system driven at its boundary by particle or energy reservoirs. Can this system be mapped, exactly or approximately, to an equilibrium system conditioned on some observable (e.g., the current)? For what class of systems and observables is this mapping possible?

\item Can we obtain the FWG theory as the low-noise limit of the driven process? In other words, for which class of processes and observables is the low-noise limit of driven process the adjoint deterministic dynamics predicted by the FWG theory?
\end{itemize}

These problems can be formulated, interestingly, in control terms. The first one, which was proposed by Evans~\cite{evans2004,evans2005a,evans2010} and which served as a direct motivation for our work on large deviation conditioning, can be rephrased by asking whether the solution of the optimal control problem of Sec.~\ref{seccont}, defined for an initial equilibrium forcing $F$, is a nonequilibrium process controlled in the stationary limit by boundary fields or forces. The second problem relates, on the other hand, to the low-noise limit of stochastic control problems and viscosity solutions of HJB equations, two problems that have been studied extensively in control theory; see \cite{fleming2006}. 

Many other problems of nonequilibrium statistical mechanics can be formulated similarly by appealing to control theory, so we expect the approach and links presented here to be useful for solving them. Of particular importance are applications to interacting systems, such as the exclusion process and the zero-range process, which have been studied recently in terms of canonical and grand-canonical versions of current conditioning in \cite{simon2009,popkov2010,popkov2011,harris2013b}. For more applications, see the references cited in Sec.~6 of \cite{chetrite2014}, and \cite{sughiyama2015}.

\subsection{Variational approximations}

The variational principles derived here are generally not easy to solve, since they involve spectral elements that are difficult to obtain analytically or numerically and require, in some cases, the determination of stationary distributions of nonequilibrium systems. However, it is possible to approximate their solutions, as commonly done in optimization theory, by restricting or \emph{projecting} the possible minimizers or maximizers on specific classes of functions.

The derivation of these approximations is also a control problem: one restricts the optimal control problem to a specific class of controllers, called a \emph{control design}. For example, one can restrict the variational principles involving the drift $u$ to drifts that are linear in the state $X_t$ or that are gradient, so as to obtain tractable approximations of large deviation functions or dominant eigenvalues. In physics, the idea of control design is equivalent to proposing \emph{ansatz}, \emph{basis functions}, \emph{trial solutions} or \emph{trial wavefunctions} to optimization problems such as the Rayleigh-Ritz principle. For applications of these approximations for jump processes and diffusions, see \cite{jack2010b,jack2014,jack2015,nemoto2011b,alexander1997,hartmann2012,hartmann2014,zhang2014}.

The quality of the approximations obtained from `limited' or `suboptimal' control designs depends on the system and observable considered, and is essentially given by the relative entropy of the `true' control solution and the `approximate' control solution obtained for a given design. Moreover, because of the maximum or minimum involved in the variational representations, one obtains not just an approximation of the true large deviation function or eigenvalue considered, but a lower bound (in the case of `sup' principles) or an upper bound (in the case of `inf' principles), which can only be improved by enlarging the class of controls considered. This monotonicity property is very useful in practice to determine the convergence and quality of approximations without the prior knowledge of `true' solutions.

\subsection{Numerical algorithms for large deviation functions}

The driven process can be used in three different ways to compute large deviation functions numerically: 
\begin{itemize}
\addtolength{\itemsep}{-5pt}
\item Implement the variational approximations described before using specific control designs or trial ansatz leading to bounds on SCGFs and rate functions. For applications of this method, see \cite{alexander1997,jack2014}.

\item Use control and dynamic programming techniques to solve the HJB equation (\ref{eqhjb1}) giving the finite-time value function $\Lambda_0^T$ and the SCGF $\Lambda_k$ in the ergodic limit. This method has not been used yet for studying time-integrated observables of nonequilibrium systems; see \cite{hartmann2012,hartmann2014,zhang2014} for related equilibrium applications in the low-temperature (FWG) limit.

\item Use the driven process as a change of measure in importance sampling.
\end{itemize}

The idea of the last method is to simulate the driven process $\hX_t$ rather than the original process $X_t$ so as to estimate the probability
\be
\Pr_{L,T}\{A_T=a\} =\Ex_{\Pr_{L,T}}[\delta(A_T-a)]
\ee
using
\be
\Pr_{L,T}\{A_T=a\} =\Ex_{\Pr_{L_{F_k},T}}\left[\delta(A_T-a) \frac{d\Pr_{L,T}}{d\Pr_{L_{F_k},T}}\right].
\label{eqis1}
\ee
The extra factor in the latter expectation is the Radon-Nikodym derivative of $X_t$ with respect to $\hX_t$ which corrects for the change of process \cite{asmussen2007}. The advantage of using $\hX_t$ for estimating (\ref{eqis1}) is that it makes the event $A_T=a$ typical for some properly chosen value $k$, as we know from the previous sections. This means that it is an efficient process for estimating large deviations: contrary to $X_t$, it does not require an exponentially large sample (in $T$) to compute the exponentially small probability $\Pr_{L,T}\{A_T=a\}$ and its corresponding rate function $I(a)$ \cite{asmussen2007,sadowsky1990,bucklew2004,touchette2011}.

That this property of the driven process can be used in practice appears questionable at first, since this process is constructed from $r_k$ and $\Lambda_k$, the very elements needed to obtain $I(a)$. Recent works have shown, however, that it is possible to construct $r_k$ and $\Lambda_k$, and therefore $\hX_t$ and $I(a)$, \emph{iteratively} or \emph{adaptively} without any prior knowledge of these functions and process by combining sampling and spectral methods. Examples of such adaptive methods have been developed by the group of Borkar \cite{borkar2004,basu2008} and by Nemoto and Sasa \cite{nemoto2014}. They fall conceptually in the class of \emph{adaptive importance sampling methods} \cite{asmussen2007}, which can be seen as a form of feedback control. There is great scope in applying these methods in physics to estimate large deviation functions, to study nonequilibrium systems, and to establish further links between these systems, rare event simulations, and control theory. 


\appendix

\section{Proofs}
\label{appproofs}

\subsection{Optimal $\rho^*$ and $j^*$}
\label{secopt1}

The minimizer $(\rho^*,j^*)$ of (\ref{eqvar1}) can be obtained by introducing Lagrange multipliers for the constraints $\nabla\cdot j=0$ and $\int \rho(x) dx=1$. We thus consider the variations
\be
\frac{\delta}{\delta \rho(a)}\left[k\tA(\rho,j)-K(\rho,j)-\int \lambda(x) \nabla\cdot j(x)\, dx-\mu\int \rho(x)\, dx\right]
\ee
and
\be
\frac{\delta}{\delta j(a)}\left[k\tA(\rho,j)-K(\rho,j)-\int \lambda(x) \nabla\cdot j(x)\, dx-\mu\int \rho(x)\, dx\right],
\ee
which yield the equations for the minimizer $(\rho^*,j^*)$
\begin{eqnarray}
0 &=&kf(a)+\frac{1}{2}(\rho^*)^{-2}[j^*(a)-J_{F,\rho^*}(a)]D^{-1}(a)[j^*(a)-J_{F,\rho^*}(a)]  \nonumber\\
& & \quad +(\rho^* D)^{-1}(a) [j^*(a)-J_{F,\rho^*}(a)]\cdot \hF(a)+\frac{1}{2}\left[\nabla\cdot \left((\rho^*)^{-1}(j^*-J_{F,\rho^*})\right)\right](a)-\mu   
\label{eqapp1}
\end{eqnarray}
and
\be
0=kg(a)-(\rho^*D)^{-1}(a) [j^*(a)-J_{F,\rho^*}(a)]+\nabla\lambda(a),
\ee
respectively. 

The second equation can be rewritten as
\begin{eqnarray}
j^*(a) & =& J_{F,\rho^*}(a)+(\rho^* D)(a) [kg(a)+\nabla\lambda(a)]\nonumber\\
&=& \left(\hF(a)+D[kg(a)+\nabla\lambda(a)]\right)\rho^*(a)-\frac{D}{2}\nabla\rho^*(a)\nonumber\\
&=&J_{F+D(kg+\nabla\lambda),\rho^*}(a)
\label{eqcc1}
\end{eqnarray}
and implies with the constraint $\nabla\cdot j^*=0$ that
\be
\rho^*=\rho^\inv_{F+D(kg+\nabla\lambda)},
\label{eqrhoc1}
\ee
which is normalized by assumption. The first equation (\ref{eqapp1}), on the other hand, can be rewritten as 
\be
\mu = \left.kf+\frac{1}{2}(kg+\nabla\lambda)D(kg+\nabla\lambda)+(kg+\nabla\lambda) \hF+\frac{1}{2}\nabla\cdot [D(kg+\nabla\lambda)]\right|_{a}.
\ee
This has the form $\mu=e^{-\lambda}(\cL_k e^{\lambda})$ with $\cL_k$ given as in (\ref{eqtiltedgen1}), so we identify $e^{\lambda}=r_k$ as the Perron-Frobenius eigenvector of $\cL_k$, since $e^{\lambda}>0$ and $\cL_k$ has only one positive eigenvector, and $\mu=\Lambda_k$ as its dominant eigenvalue. From equations (\ref{eqcc1}) and (\ref{eqrhoc1}) and the definition (\ref{eqmoddrift1}) of the modified drift $F_k$, we then obtain
\be
\rho^*=\rho_{F_k}^\inv,\qquad j^*=J_{F_k,\rho_{F_k}^\inv}
\ee
with $\rho_{F_k}^\inv=r_kl_k$.

\subsection{Donsker-Varadhan result for $g=0$}
\label{appdv}

In the case $g=0$, the variational representation (\ref{eqvar1}) reduces by direct minimization on $j$ to the DV variational principle \cite{donsker1975b}, which is in our notations
\be
\Lambda_k=\sup_{\rho:\int \rho(x) dx=1} \left\{\int kf(x)\rho(x)\, dx- I_2(\rho) \right\},
\label{eqdvr1}
\ee
where
\be
I_2(\rho)=\inf_j I(\rho,j)= -\inf_{h>0} \int \rho(x) (h^{-1} Lh )(x)\, dx
\label{eql2rf1}
\ee
is the rate function of the empirical density $\rho_T$ for the Markov process with generator $L$, first derived by Donsker and Varadhan and now often referred to as the \emph{level-2 rate function} \cite{donsker1975,donsker1975a,donsker1976}. 

This rate function is known to be explicitly given by 
\be
I_2(\rho)=-\int dx\, \rho_F^\inv(x) \sqrt{\frac{\rho(x)}{\rho_F^\inv(x)}} \left(L \sqrt{\frac{\rho}{\rho_F^\inv}}\right)(x)
\label{eqrevrf1}
\ee
when $X_t$ is reversible with respect to the invariant density $\rho^\inv_F$, that  is, $\rho_F^\inv L(\rho_F^\inv)^{-1}=L^\dag$. As a particular case, if $L$ is hermitian, that is, if $X_t$ is reversible with respect to the constant (Lebesgue) density, then (\ref{eqdvr1}) reduces to
\begin{eqnarray}
\Lambda_k &=& \sup_{\rho: \int \rho(x) dx=1}\left\{ \int dx\, \left[k f(x)\rho(x) +\rho^{1/2}(x)  (L\rho^{1/2})(x)\right]\right\}\nonumber\\
& =& \sup_{\sigma: \int\sigma^2(x)dx=1}\left\{ \int dx\, \left[kf(x)\sigma^2(x)+\sigma(x) (L\sigma)(x)\right]\right\}\nonumber\\
&=& \sup_{\sigma: \int\sigma^2(x)dx=1}\left\{ \int dx\, \sigma(x) (\cL_k\sigma)(x)\right\},
\end{eqnarray}
which is the Rayleigh-Ritz variational principle for $\cL_k$ \cite{courant2004,kac1951}.

\subsection{Relative entropy representations via level-2.5 large deviations}
\label{appreproof1}

The following representation of the level-2.5 rate function can be derived from a result of Barato and Chetrite \cite{barato2015}: 
\be
K(\rho,j)=
\left\{
\begin{array}{lll}
\displaystyle\lim_{T\ra\infty}\frac{1}{T}S(\Pr_{L_{\tu},T}|| \Pr_{L,T}) & & \text{if } \nabla\cdot j=0\\
\infty & & \text{otherwise,}
\end{array}
\right.
\label{eqrelrf1}
\ee
where $\Pr_{L_{\tu},T}$ is the path measure of a modified diffusion with drift $\tu$, chosen in such a way that the typical behavior of $(\rho_T,j_T)$ in the $T\ra\infty$ limit is $(\rho,j)$; that is, $\tu$ is such that
\be
(\rho_T,j_T)\Pra{\Pr_{L_{\tu},T}} (\rho_{\tu}^\inv,J_{\tu,\rho_{\tu}^\inv})= (\rho,j)
\label{equlim1}
\ee
or explicitly by (\ref{eqcurr1}),
\be
\tu = \frac{j}{\rho}+\frac{D}{2}\nabla \ln \rho.
\ee
With the change of variables $(\rho,j)\ra (\rho,u)$ with $J_{u,\rho}=j$ introduced before in (\ref{eqvar2}), we then get $\tu=u$ and
\be
K(\rho_u^\inv,J_{u,\rho_u^\inv})=\lim_{T\ra\infty}\frac{1}{T}S(\Pr_{L_u,T}||\Pr_{L,T})
\label{eqrelurf1}
\ee
for the function $K(\rho_u^\inv,J_{u,\rho_u^\inv})$ shown in (\ref{eqrfu1}).

The representation (\ref{eqrelrf1}) can be substituted directly into (\ref{eqvar1}) and (\ref{eqvarfct1}) to obtain representations for $\Lambda_k$ and $I(a)$, respectively, involving $(\rho,j)$ and the drift $\tu$ defined in (\ref{equlim1}). Similarly, we can substitute the more explicit formula (\ref{eqrelurf1}) above into (\ref{eqvar2}) and (\ref{eqvarfct2}) to obtain relative entropy representations of $\Lambda_k$ and $I(a)$, respectively, which are now explicit in $u$. From these, we obtain the final representations (\ref{eqrerep1}) and (\ref{eqrerep2}) announced in Sec.~\ref{secre1} by noting that the limit in probability (\ref{equlim1}) also implies a limit in mean, so that
\be
\tA(\rho_u^\inv,J_{u,\rho_u^\inv})=\lim_{T\ra\infty} \Ex_{\Pr_{L_u,T}}[A_T]=\lim_{T\ra\infty} \Ex_{\Pr_{L_u,T}}[\tA(\rho_T,j_T)].
\ee

From this proof, it is clear that the variational representations of $\Lambda_k$ and $I(a)$ expressed via $(\rho^*,j^*)$ and $u$ are equivalent to the relative entropy representations of these functions. It is also clear that the solution $(\rho^*,j^*)$ of the Laplace principle (\ref{eqvar1}) or the contraction principle (\ref{eqvarfct1}) is the typical value of $(\rho_T,J_T)$ in the driven process, as noted before in (\ref{eqminsol1}): the solution of (\ref{eqrerep1}) or (\ref{eqrerep2}) is $u^*=F_k$, so that $\tu(\rho^*,j^*)=u^*$, implying (\ref{eqminsol1}) from (\ref{equlim1}). 

\subsection{Eigenfunction representation (\ref{eqrepse1})}
\label{appserep1}

The modified variational principle (\ref{eqrepse1}) follows from the variational principle (\ref{eqvar2}) involving the drift $u$ by performing the contraction $u\ra h$ with
\be
u=F+D(kg+\nabla\ln h). 
\label{eqhdrift1}
\ee
Since the maximum of (\ref{eqvar2}) has this form, we do not restrict this result by rewriting it with (\ref{eqcurr1}) as
\be
\Lambda_k
 =  \sup_{h>0} \int dx\, \rho^\inv_{F+D(kg+\nabla\ln h)} \left[ kf+kg\cdot \hF +\frac{k}{2}\nabla\cdot (Dg)+\frac{1}{2}(kg+\nabla\ln h)D(kg-\nabla\ln h)
\right].\qquad
\label{eqintr1}
\ee
The term in brackets can be expressed as
\be
kf+kg\cdot \hF +\frac{k}{2}\nabla\cdot (Dg)+\frac{1}{2}(kg+\nabla\ln h)D(kg-\nabla\ln h)=h^{-1}(\cL_k h)-(\cL_k^h \ln h),
\label{eqintr2}
\ee
where 
\be
\cL_k^h = h^{-1} \cL_k h-h^{-1}(\cL_k h)
\ee
is the \emph{generalized Doob $h$-transform} of $\cL_k$ \cite{chetrite2014}. Similarly to (\ref{eqgendoob1}), this transform defines a new diffusion with drift $u$ given by (\ref{eqhdrift1}) and stationary density $\rho^\inv_u$. Consequently,
\be
\int dx\, \rho^\inv_{F+D(kg+\nabla\ln h)}\, (\cL_k^h \ln h)=\int dx\, \left(\cL_k^{h\dag} \rho^\inv_{F+D(kg+\nabla\ln h)}\right) \ln h=0,
\ee
so that (\ref{eqintr1}) combined with (\ref{eqintr2}) reduces to (\ref{eqrepse1}).

\subsection{Eigenfunction representations (\ref{eqrepse2}) and (\ref{eqrepse3})}
\label{appserep2} 

The representation (\ref{eqrepse3}) for the rate function can be derived from the representation (\ref{eqvarfct1}) involving $(\rho,j)$. In the case $g=0$, 
\be
\tA(\rho,j)=\int dx\, \rho(x)f(x),
\ee
so that (\ref{eqvarfct1}) becomes
\be
I(a)=\inf_{\rho:\topcons{\int \rho(x)dx=1}{\int \rho(x)f(x)dx=a}} \inf_j K(\rho,j)=\inf_{\rho:\topcons{\int \rho(x)dx=1}{\int \rho(x)f(x)dx=a}} I_2(\rho),
\ee
where $I_2(\rho)$ is the level-2 rate function. From the expression of this function shown in (\ref{eql2rf1}), we thus get
\be
I(a)=\inf_{\rho:\topcons{\int \rho(x)dx=1}{\int \rho(x)f(x)dx=a}} \left\{-\inf_{h>0} \int \rho(x) (h^{-1} Lh)(x)\right\}.
\label{eqedv1}
\ee
To obtain (\ref{eqrepse3}), we then only need to perform the following change of variables:
\be
(h,\rho)\ra (r,l)=(h,\rho h^{-1}).
\ee
It is not clear whether this change of variables preserves the minimal nature of $h$ in (\ref{eqedv1}); hence the transformation of the infimum in (\ref{eqedv1}) to a stationary point optimization in (\ref{eqrepse3}).

The same argument applied to (\ref{eqvar1}) yields (\ref{eqrepse2}) knowing that $\cL_k=L+kf$ for $g=0$. In both cases, the solution $(r^*,l^*)=(r_k,l_k)$, or equivalently $(h^*,\rho^*)=(r_k,r_kl_k)$, is obtained by solving the infimum on $h$ in (\ref{eqedv1}) for the known solution $\rho^*=r_kl_k$.

\subsection{Modified HJB equation for current costs}
\label{apphjb}

We want to derive the HJB equation for the stochastic control problem on the finite horizon $[t,T]$ involving the controlled diffusion $X_t^u$ defined in (\ref{eqdiff1}) and the following cost function:
\be
C_t^T(x)=\inf_u \Ex_{\Pr_{L_u,T}}\left[\int_t^T \psi(X_s^u,u_s)ds+\phi(X^u_s)\circ dX_s^u\right],
\ee
where the expectation is with respect to the law of the controlled process started at $X_t^u=x$. The part involving $\psi$ is the usual cost considered in optimal control theory; the added current or Stratonovich cost involving $\phi$ has been considered only recently by Chernyak, Chertkov, Bierkens and Kappen~\cite{chernyak2014} who give a partial solution for quadratic costs in the control drift $u_t$. 

More general results can be obtained in a very simple way by showing that the term $\phi$ can be absorbed in $\psi$ to rewrite $C_t^T$ as
\be
C_t^T(x)=\inf_u \Ex_{\Pr_{L_u,T}}\left[\int_t^T \hpsi(X_s^u,u_s)ds\right]
\label{eqmcf1}
\ee
with the modified cost
\be
\hpsi(x,u)=\psi(x,u)+\phi(x)\cdot \hat u(x)+\frac{1}{2}\nabla\cdot (D\phi(x)).
\label{eqmc1}
\ee
This follows by noting that
\begin{multline}
\qquad\Ex_{\Pr_{L_u,T}}\left[\int_t^T \psi(X_s^u,u_s)ds +\phi(X^u_s)\circ dX_s^u\right]=\\
 \int_t^T \int dy\, \left[\psi(y,u_s)(P_{L_u})_t^s(x,y)\, ds+\phi(y)\cdot J_{u,(P_{L_u})_t^s(x,\cdot)}(y) \right],
\label{eqexpcost1}
\end{multline}
where
\be
(P_{L_u})_s^t(x,y)=e^{(s-t)L_u}(x,y)
\ee 
is the transition probability for the controlled diffusion between $X_t=x$ and $X_s=y$ with $s>t$, and $J_{u,(P_{L_u})_t^s(x,\cdot)}(y)$ is the associated  probability current defined in (\ref{eqcurr1}). Inserting this definition in (\ref{eqexpcost1}) and using integration by parts yields $\hpsi$ as above.

In the purely additive form involving in $\hpsi$, the cost $C_t^T(x)$ now satisfies the usual backward HJB equation
\be
-\p_t C_t^T(x)=\inf_u \{ \hpsi(x,u)+L_u C_t^T(x)\}
\ee
with $C_T^T=0$. Given (\ref{eqmc1}), we therefore obtain
\be
-\p_t C_t^T(x)=\inf_u \left\{ \psi(x,u)+\phi(x)\cdot \hat u(x)+\frac{1}{2}\nabla\cdot (D\phi(x))+L_u C_t^T(x)\right\}.
\label{eqgenhjb1}
\ee
It can be checked that this recovers the results of \cite{chernyak2014} for quadratic costs. 

\subsection{Control representation}
\label{appcontrep}

The PDE satisfied by the exponential cost $G_t^T(x,k)=e^{\Lambda_t^T(x,k)}$ is obtained by inserting the solution (\ref{eqver1}) for $u^*$ in the HJB equation (\ref{eqhjb1}) and by using the expression of the generator $L_u$:
\begin{eqnarray}
-\frac{\p_tG_t^T}{G_t^T} & = & kf+\frac{k}{2}\nabla\cdot(Dg)+\frac{1}{2}\left(kg+\frac{\nabla G_t^T}{G_t^T}\right)D\left(kg+\frac{\nabla G_t^T}{G_t^T}\right)
 +\hF\cdot\left(kg+\frac{\nabla G_t^T}{G_t^T}\right)+\frac{1}{2}\nabla\cdot\frac{D\nabla G_t^T}{G_t^T}\nonumber\\
 & = & kf+\frac{k}{2}\nabla\cdot (Dg)-\frac{k}{2}(Dg)\cdot\left(kg+2\frac{\nabla G_t^T}{G_t^T}\right)
+\hF\cdot\left(kg+\frac{\nabla G_t^T}{G_t^T}\right)+\frac{1}{2}\nabla\cdot\frac{D\nabla G_t^T}{G_t^T}.
 \end{eqnarray}
In these equations, all gradients are in $x$. Consequently,
\be
\p_t G_t^T+k\left(f+\frac{1}{2}\nabla\cdot (Dg)+\frac{k}{2}gDg+\hF\cdot g\right)G_t^T+\left(\hF+kDg\right)\cdot\nabla G_t^T+\frac{1}{2}\nabla\cdot\frac{D\nabla G_t^T}{G_t^T}=0,
\ee
which can be rewritten with the tilted generator $\cL_k$ (\ref{eqtiltedgen1}) as
\be
(\p_t +\cL_k) G_t^T=0,\qquad G_{T}^{T}=1,
\ee
as claimed in (\ref{eqhjb2}).

\section{Jump processes}
\label{appjp}

We translate in this section the results of the previous sections in the language of pure jump processes. The notations follow as before those of \cite{chetrite2014}. 

We consider an ergodic pure jump process $X_t$, $t\in[0,T]$, defined by the \emph{transition kernel} $W(x,y)$ representing the transition rate (probability density per unit time) for the transition going from $x$ to $y$. The generator $L$ of this process is expressed in terms of $W(x,y)$ as
\be
Lh(x)=\int dy\, W(x,y)[h(y)-h(x)],
\ee
where $h(x)$ is a bounded measurable function on the space of $X_t$.\footnote{Integrals must be replaced by sums in this appendix if $X_t$ lives in a discrete space.} We also define from $W(x,y)$ the \emph{escape rates}
\be
\lambda(x)=\int W(x,y)\, dy =(W1)(x).
\ee
With these elements, it is then common to express the generator as $L=W-\lambda$.

For jump processes, the general observable $A_T$ defined in (\ref{eqobs1}) must be modified to account for the fact that paths of these processes have discontinuities. This leads us to consider
\be
A_T=\frac{1}{T}\int_0^T f(X_t)dt+\frac{1}{T}\sum_{t:\Delta X_t\neq 0} g(X_{t^-},X_{t^+}),
\ee
where the sum is over all times $t$ at which a jump occurs with state $X_{t^-}$ before the jump and $X_{t^+}$ after the jump. The choice of functions $g(x,y)$ and $f(x)$ depends on the application or physical observable considered. Choosing $f=0$ and $g(x,y)=1$, for example, gives the number of jumps per unit time occurring in $[0,T]$, which is called the \emph{activity} \cite{lecomte2005,lecomte2007,maes2008a}, while $f=0$ and $g(x,y)=-g(y,x)=1$ gives the current per unit time \cite{maes2008a,barato2015}. 

The large deviations of $A_T$ can be determined similarly as for diffusions from the SCGF $\Lambda_k$, which corresponds to the largest eigenvalue of the tilted generator
\be
\cL_k=We^{kg}-\lambda+kf,
\ee
where the first term on the right-hand side of the equation is understood as the Hadamard component-wise product \cite{chetrite2014}. As before, we denote the eigenfunction corresponding to $\Lambda_k$ by $r_k$ and the dual eigenfunction by $l_k$. The driven process associated with the conditioned process $X_t|A_T=a$ is then defined exactly as in (\ref{eqgendoob1}) by a generalized Doob transform of $\cL_k$, which yields for jump processes the driven generator $L_k=W_k-\lambda_k$ involving the driven rates
\be
W_k(x,y)=r_k^{-1}(x) W(x,y)e^{kg(x,y)}r_k(y)
\ee
and the driven escape rates $\lambda_k(x)=(W_k1)(x)$ \cite{chetrite2014}.

The equivalence results expressed by (\ref{eqproceq1}) or (\ref{eqeq1}) also hold with the path measure $\Pr_{L_{F_k},T}$ replaced by the path measure $\Pr_{W_k,T}$ of the jump process with rates $W_k$ and imply, similarly to the diffusion case, that the driven jump process $\hX_t$ is equivalent to the conditioned jump process $X_t|A_T=a$ at the level of stationary states. In particular, both processes have the same empirical density $\rho_T(x)$ in the limit $T\ra\infty$, which converges to $\rho^\inv_{W_k}(x)=r_k(x)l_k(x)$. They also have the same asymptotic empirical current, which in the case of jump processes is defined as
\be
J_T(x,y)=C_T(x,y)-C_T(y,x),
\ee
where
\be
C_T(x,y)=\frac{1}{T}\sum_{t:\Delta X_t\neq 0} \delta (X_{t^-}-x)\delta(X_{t^+}-y)
\ee
is the so-called \emph{empirical flow} \cite{bertini2012} corresponding to the number (per unit time) of jumps from $x$ to $y$. The latter quantity converges in the driven jump process to 
\be
C_{W_k,\rho^\inv_{W_k}}(x,y)=\rho_{W_k}^\inv(x) W_k(x,y).
\label{eqef1}
\ee
Therefore,
\be
J_T\Pra{\Pr_{W_k,T}} J_{W_k,\rho^\inv_{W_k}}(x,y)=\rho_{W_k}^\inv(x)W_k(x,y)-W_k(y,x)\rho_{W_k}^\inv(y).
\ee
For $I(a)$ convex, we then also have
\be
\rho_T\Pra{\Pr_{a,T}^\micro}\rho_{W_k}^\inv,\qquad C_T\Pra{\Pr_{a,T}^\micro} C_{W_k,\rho_{W_k}^\inv},\qquad J_T\Pra{\Pr_{a,T}^\micro} J_{W_k,\rho^\inv_{W_k}}
\ee
in conditioned jump process $X_t|A_T=a$ for $k=I'(a)$.

These results are essentially the same as for diffusions, except for the definition of the empirical current. For building the contraction of $A_T$, we also need the pair $(\rho_T,C_T)$ rather than $(\rho_T,J_T)$. As a function of $\rho_T$ and $C_T$, we indeed have
\be
\tA(\rho_T,C_T)=\int f(x) \rho_T(x) dx+\int g(x,y) C_T(x,y)\, dx dy.
\ee
Moreover, $(\rho_T,C_T)$ is known to satisfy an LDP with rate function
\be
K(\rho,C)=\int dx dy \left(C(x,y)\ln\frac{C(x,y)}{\rho(x)W(x,y)}-C(x,y)+\rho(x)W(x,y)\right)
\label{eqrfjp1}
\ee
if
\be
\int C(x,y) dy=\int C(y,x)dy
\label{eqbal1}
\ee 
for all $x$ and $I(\rho,C)=\infty$ otherwise. More detail about these results can be found in \cite{maes2008a,bertini2012,barato2015}.

From here, we translate our results of the previous sections as follows. 

First, we obtain a jump version of the first variational representation (\ref{eqvar1}) using $B_T=(\rho_T,C_T)$ and the rate function $K(\rho,C)$ in (\ref{eqrfjp1}) to get
\be
\Lambda_k=\sup_{\rho,C}\{k\tA(\rho,C)-K(\rho,C)\}.
\label{eqscgfjp1}
\ee
It is understood that the minimization is over all densities $\rho$ such that $\int \rho(x)dx=1$ and balanced flows $C$ satisfying (\ref{eqbal1}). The minimizer, as expected and as can be checked explicitly, is $\rho^*=\rho_{W_k}^\inv$ and $C^*(x,y)=\rho^\inv_{W_k}(x)W_k(x,y)$.

Second, we can re-parameterize the minimization in (\ref{eqscgfjp1}) in terms of a transition matrix that determines the ergodic limit of both $\rho_T$ and $C_T$ to obtain jump version of the variational representation (\ref{eqvar2}). To be more precise, let us denote by $\rho_Q^\inv$ the invariant density of an ergodic jump process with transition rate matrix $Q(x,y)$, and let $C_{Q,\rho_Q^\inv}$ be the invariant empirical flow (\ref{eqef1}) obtained with the same transition matrix. Given the change of variables $C\ra Q$ with $C=C_{Q,\rho}$, the constraint (\ref{eqbal1}) then implies $\rho=\rho_Q^\inv$, so that
\be
\Lambda_k=\sup_{Q} \left\{ \tA(\rho_Q^\inv,C_{Q,\rho_Q^\inv})-K(\rho_Q^\inv,C_{Q,\rho_Q^\inv})\right\}, 
\label{eqjprep2}
\ee
where
\be
K(\rho_Q^\inv,C_{Q,\rho_Q^\inv})=\int dxdy\, \rho_Q^\inv(x)\left[ Q(x,y)\ln\frac{Q(x,y)}{W(x,y)}-Q(x,y)+W(x,y)\right]
\label{eqcontcost1}
\ee
is the level-2.5 rate function (\ref{eqrfjp1}) expressed with the flow (\ref{eqef1}) and transition matrix $Q$. In this form, it can be checked that the minimizer $Q^*$ is $W_k$, the transition rate matrix of the driven jump process.

Third, the representation (\ref{eqvarfct1}) of the rate function $I(a)$ becomes by considering the constrained maximization (\ref{eqcp2}) with $B_T=(\rho_T,C_T)$ and the rate function $K(\rho,C)$:
\be
I(a)=\inf_{(\rho,C):\tA(\rho,C)=a} K(\rho,C),
\ee
which can be re-written as
\be
I(a)=\inf_{Q:\tA(\rho_Q^\inv,C_{Q,\rho_Q^\inv})=a} K(\rho_Q^\inv,C_{Q,\rho^\inv_Q})
\ee
using the same re-parameterization as before.

Fourth and finally, we can consider the jump process $X_t^Q$ as being controlled by a choice of transition rates $Q(x,y)$ to rewrite all these representations in control form. The main result worth noting in this case is
\be
\Lambda_k=\lim_{T\ra\infty}\sup_Q \{kA_T -K_T\}
\label{eqcontjp1}
\ee
for almost all paths, where 
\be
A_T=\frac{1}{T}\int_0^T f(X_t^Q)dt+\frac{1}{T}\sum_{t:\Delta X_t^Q\neq 0} g(X_{t^-}^Q,X_{t^+}^Q)
\ee
is the value of the observable obtained with respect to the \emph{$Q$-controlled process} $X_t^Q$ and 
\be
K_T=\frac{1}{T}\int_0^T \int dy [Q(X_t^Q,y)-W(X_t^Q,y)]+\frac{1}{T}\sum_{t:\Delta X_t^Q\neq 0} \ln\frac{Q(X_{t^-}^Q,X_{t^+}^Q)}{W(X_{t^-}^Q,X_{t^+}^Q)}
\label{eqktjp1}
\ee
is the empirical version of the rate function shown in (\ref{eqcontcost1}). Equation (\ref{eqcontjp1}) is the jump analog of our control result (\ref{eqcontd1}). It was  obtained before in mean form by Jack and Sollich \cite{jack2015}.

We do not translate the variational representations involving the relative entropy, since they are obtained similarly as for diffusions. These representations rationalize in terms of large deviation functions the results derived for jump processes by Monthus \cite{monthus2011}, who refers to the limit of the path relative entropy as the \emph{Kolmogorov-Sinai entropy}. Some connections that exist between the driven process, the maximum caliber method, and the effective transition rates introduced by Evans \cite{evans2004,evans2005a,evans2010} (see \cite{chetrite2014} for more details) are also mentioned in \cite{monthus2011}.

\section{Markov chains}
\label{appmc}

The translation of our results for an ergodic Markov chain $\{X_i\}_{i=0}^{N}$ with transition matrix $M(x,y)$ closely follows the previous appendix and the Appendix E of \cite{chetrite2014}, so we shall be brief in this section. The quantities and concepts to consider are as follows:
\begin{itemize}
\item Observable:
\be
A_{N}=\frac{1}{N}\sum_{i=0}^{N-1} g(X_i,X_{i+1}),
\label{eqobsmc1}
\ee
where $g$ is an arbitrary function. This observable includes one-point observables by choosing $g(x,y)=f(x)$.

\item Tilted matrix: The SCGF $\Lambda_k$ is now the logarithm of the dominant eigenvalue of the matrix
\be
\cM_k(x,y)=M(x,y)e^{k g(x,y)}.
\ee
As before, we denote by $r_k$ the associated eigenvector of $\cM_k$ and by $l_k$ the eigenvector of the dual (transpose) of $\cM_k$, which is nothing but the left eigenvector of $\cM_k$.

\item Driven Markov chain: The discrete-time version of the driven process is the Markov chain with modified transition probabilities
\be
M_k(x,y)=r^{-1}_k(x) \cM_k(x,y)r_k(y) \, e^{-\Lambda_k}.
\ee
The stationary density of this process is also $\rho_{M_k}^\inv(x)=r_k(x)l_k(x)$; see Appendix E of \cite{chetrite2014}.

\item Empirical density:
\be
\rho_{1,N}(x)=\frac{1}{N}\sum_{i=0}^{N-1} \delta_{X_i,x},
\ee
where $\delta_{x,y}$ is the Kronecker symbol.

\item Pair empirical density:
\be
\rho_{2,N}(x,y)=\frac{1}{N}\sum_{i=0}^{N-1}\delta_{x,X_i}\delta_{y,X_{i+1}}.
\ee
This is the Markov chain analog of the empirical flow $C_T(x,y)$.

\item Contraction: The empirical density is not needed to obtain a representation of $A_N$, as defined in (\ref{eqobsmc1}); the pair empirical density is sufficient:
\be
\tA(\rho_2)=\int g(x,y) \rho_2(x,y)\, dxdy.
\label{eqconfmc1}
\ee

\item LDP for the representing observable: The pair empirical density is known to satisfy an LDP with rate function \cite{dembo1998,hollander2000,touchette2009}
\be
K(\rho_2)=
\left\{
\begin{array}{lll}
\displaystyle\int dxdy\, \rho_2(x,y)\ln \frac{\rho_2(x,y)}{\int dz \rho_2(x,z) M(x,y)} & & \text{if }\int \rho_2(x,y)dx=\int \rho_2(y,x)dx\\
\infty & & \text{otherwise}.
\end{array}
\right.
\label{eqrfmc1} 
\ee

\item Typical asymptotic states in the driven Markov chain: $\rho_{2,N}(x,y)$ converges in the limit $N\ra\infty$ to the stationary joint distribution of the ergodic Markov chain considered. For the driven process, we thus have
\be
\rho_{2,N}(x,y)\Pra{\Pr_{M_k,N}} \rho_{M_k}^\inv(x) M_k(x,y)=l_k(x) \cM_k(x,y) r_k(y)\, e^{-\Lambda_k}.
\ee
By contraction, this also implies
\be
\rho_{1,N}(x)\Pra{\Pr_{M_k,N}} \rho_{M_k}^\inv(x)
\ee
for the driven Markov chain.

\item Equivalence with conditioned Markov chain: Assuming that $A_N$ satisfies an LDP with convex rate function $I(a)$, we have the same two limits above for the conditioned Markov chain $X_n|A_N=a$ with $k=I'(a)$. 
\end{itemize}

The results of the previous sections are expressed in terms of these notations with minor changes. We only note the variational formula
\be
\Lambda_k=\sup_{\rho_2}\{k\tA(\rho_2)-K(\rho_2)\},
\label{eqrepmc1}
\ee
which derives from the Laplace principle (\ref{eqcp1}), its transition matrix version
\be
\Lambda_k=\sup_{Q}\{k\tA(\rho_Q^\inv\otimes Q)-K(\rho_Q^\inv\otimes Q)\},
\label{eqrepmc2}
\ee
and the control representation
\be
\Lambda_k=\lim_{N\ra\infty}\sup_Q \{kA_N-K_N\}.
\label{eqrepmc3}
\ee
The last variational representation involves the observable
\be
A_N=\frac{1}{N}\sum_{i=0}^{N-1} g(X_i^Q,X_{i+1}^Q)
\ee
accumulated by the \emph{controlled Markov chain} $\{X_i^Q\}_{i=0}^{N}$ with transition matrix $Q(x,y)$ and the empirical version of the relative entropy:
\be
K_N=\frac{1}{N}\sum_{i=0}^{N-1}\ln \frac{P(X_i^Q,X_{i+1}^Q)}{M(X_i^Q,X_{i+1}^Q)}.
\ee
The representation (\ref{eqrepmc1}) is the Markov chain analog of (\ref{eqvar1}) and (\ref{eqscgfjp1}), while (\ref{eqrepmc2}) is the Markov chain analog of (\ref{eqvar2}) and (\ref{eqjprep2}). The latter result, expressed explicitly with the contraction (\ref{eqconfmc1}) and rate function (\ref{eqrfmc1}), was previously derived by Sasa \cite{sasa2012}. 

Markov chain versions of the representations involving the rate function and the relative entropy follow similarly. As for the jump process case, they rationalize from the large deviation point of view the results of Monthus \cite{monthus2011}.

\begin{acknowledgments}

We thank Florian Angeletti, Patrick Cattiaux, Krzysztof Gawedzki, Vivien Lecomte, and Christian L\'eonard for useful discussions. We are also grateful for the hospitality and support of the Galileo Galilei Institute for Theoretical Physics and INFN during the workshop `Advances in Nonequilibrium Statistical Mechanics', and the Universit\`a di Modena e Reggio Emilia where this work was finished. Further financial support was provided by the ANR STOSYMAP (ANR-2011-BS01- 015), the National Research Foundation (NRF) of South Africa (project CSUR 13090934303), NITheP (visitors programme), and Stellenbosch University (project funding for new appointee).
\end{acknowledgments}

\bibliography{masterbib}

\end{document}